\documentclass[11pt,a4paper]{article}
\pdfoutput=1

\usepackage{epsfig,latexsym,amsfonts,amsmath,amsthm,amssymb,amsbsy,multirow,slashed,color,mathrsfs,subfigure,wrapfig,comment}
\usepackage[font={footnotesize,sl},bf]{caption}
\usepackage{jheppub}

\newcommand\be{\begin{equation}}
\newcommand\bea{\begin{eqnarray}}
\newcommand\ee{\end{equation}}
\newcommand\eea{\end{eqnarray}}

\newcommand{\bdm}{\begin{displaymath}}
\newcommand{\edm}{\end{displaymath}}
\newcommand{\nn}{\nonumber \\}
\newcommand{\f}[2]{\frac{#1}{#2}}
\newcommand{\bref}[1]{(\ref{#1})}

\newcommand{\ket}[1]{|#1 \rangle}
\newcommand{\bra}[1]{\langle #1 |}

\newcommand{\U}{\mathcal{U}}
\newcommand{\too}{\longrightarrow}

\usepackage{color}

\newcommand{\braket}[1]{\langle #1 \rangle}

\title{Unitarity and fuzzball complementarity: \\ ``Alice fuzzes but may not even know it!"}
\author[1]{Steven G.\ Avery,}
\author[2]{Borun D.\ Chowdhury,}
\author[3]{Andrea Puhm}
\affiliation[1]
{The Institute of Mathematical Sciences,\\
CIT Campus, Taramani, Chennai, India 600113}
\affiliation[2]
{Institute for Theoretical Physics, University of Amsterdam,\\
Science Park 904, Postbus 94485, 1090 GL Amsterdam, The Netherlands}
\affiliation[3]
{Institut de Physique Th\'eorique, CEA Saclay, 91191 Gif sur Yvette, France}

\abstract{
We investigate the recent black hole firewall argument. For a black hole in a typical state we argue that unitarity requires every quantum of radiation leaving the black hole to carry information about the initial state. An information-free horizon is thus inconsistent with unitary at every step of the evaporation process (in particular {\em both} before and after Page time). The required horizon-scale structure is manifest in the fuzzball proposal which provides a mechanism for holding up the structure. In this context we want to address the experience of an infalling observer and discuss the recent \emph{fuzzball complementarity} proposal.
Unlike black hole complementarity and observer complementarity which postulate asymptotic observers experience a hot membrane while infalling ones pass freely through the horizon, fuzzball complementarity postulates that fine-grained operators experience the details of the fuzzball microstate and coarse-grained operators experience the black hole. In particular, this implies that an infalling detector tuned to energy $E \sim T_H$, where $T_H$ is the asymptotic Hawking temperature, does not experience free infall while one tuned to $E \gg T_H$ does.
}

\begin{document}
\maketitle

\section{Introduction} \label{Introduction}

General relativity predicts that for a sufficiently large quantity of matter in a fixed spatial region, gravitational interactions dominate over all others and the matter inevitably collapses to form a black hole, cf.~\cite{Gibbons:2009xm}. 
Regardless of the matter's initial configuration, the black hole that finally results is essentially unique with only a few independent parameters. Furthermore, if one considers quantum fields in the background of the classical black hole solution, one finds that 
unlike for an ordinary star, a black hole's ``surface", the event horizon, is in a unique state independent of the collapsing matter.
This state, the Unruh vacuum, allows an infalling observer to fall freely through it.  On much longer time scales, the black hole evaporates away as pairs of particles are pulled out of the Unruh vacuum at the horizon.
Since the horizon contains no information of the original matter, the created pairs are in a special state independent of the matter's initial state. Hawking's calculation~\cite{Hawking:1974sw} implies that the resulting radiation is in a mixed state and therefore the evaporation process  is manifestly non-unitary. This presents a foundational conflict between general relativity and quantum mechanics that any complete theory of quantum gravity must resolve. 

The principle of black hole complementarity (BHC) \cite{Susskind:1993if} was advanced to reconcile the ideas of unitary evaporation of black holes and free infall through the horizon of the black hole. BHC postulates that, as far as an asymptotic observer (one who stays outside the black hole forever) is concerned, the black hole behaves like a hot membrane emitting Hawking-like quanta in a unitary fashion; however, BHC also certifies that an infalling observer does not crash into this membrane but instead experiences the infalling vacuum state (Unruh vacuum) at the horizon and therefore nothing unusual until the singularity is approached. In BHC not only is the microscopic description of the membrane, i.e. the {\em mechanism} for information retrieval, missing, the idea of free infall is in considerable tension with unitarity of the evaporation process. The latter was made sharp in a recent Gedankenexperiment of AMPS~\cite{Almheiri:2012rt} and discussed further in \cite{Bousso:2012as,Mathur:2012jk,Chowdhury:2012vd,Susskind:2012rm,Banks:2012nn,Ori:2012jx,Susskind:2012uw,Hossenfelder:2012mr}.
Because of this tension it is useful to focus on unitarity and infall questions successively: 

\begin{itemize}
\item {\bf Unitarity}

To address the problem of unitarity Mathur proposed the fuzzball conjecture. This proposal states that the true microstates accounting for the Bekenstein--Hawking entropy are horizonless and singularity-free ``fuzzballs''. According to this proposal the radiation emanates locally from the fuzzballs, instead of from an information-free horizon, and thus there is no information loss.


The fuzzball proposal is a huge departure from the conventional black hole picture in that it basically states that a black hole is a complicated quantum star of horizon-scale size. Many have held the view that giving up the traditional horizon is not necessary to save unitarity and (erroneously) believe that, since the number of emitted quanta is very large $(O(M^2))$,\footnote{In this paper we work in Planck units.} small corrections to the leading order Hawking process, for which the semi-classical physics with the horizon being approximately the vacuum state remains trustworthy, can accumulate over time and restore unitarity.

This hope is based on Page's argument \cite{Page:1993df} that information of complex evaporating systems is encoded in delicate correlations between radiation quanta.  Concretely, Page has shown that for a macroscopically large evaporating system in a typical state where all subsystems are almost maximally entangled very little information can be read off unless one can access more than half of the system.  From this one concludes that the von Neumann entropy of the radiation emitted from the burning system grows linearly, then turns around at the ``Page time'' where half of the system has evaporated and then falls linearly to zero as shown in Figure~\ref{evaporation}a~\cite{Page:1993up}.  If black hole evaporation were unitary then a black hole in a typical state would evaporate according to Figure~\ref{evaporation}a.  In a recent series of papers~\cite{Mathur:2009hf,Mathur:2010kx,Mathur:2011wg,Mathur:2011uj,Mathur:2012np}, however, Mathur showed that for black hole evaporation with only {\em small} 
corrections to the leading-order Hawking process the von Neumann entropy of radiation keeps monotonically increasing in time with no turnover, as shown in Figure~\ref{evaporation}b. To get an evolution as in Figure~\ref{evaporation}a, he therefore argues that the small correction have to be made large and we can no longer trust effective field theory at the horizon scale.\footnote{Unfortunately, the explicit small correction considered in~\cite{Mathur:2009hf} when made arbitrarily large cannot make the von Neumann entropy of the radiation turn over~\cite{Giddings:2012dh}. The quantitative argument in~\cite{Mathur:2009hf} is thus unconvincing; however, one of the authors~\cite{Avery:2011nb} showed that arbitrary small corrections (including ones that when made large restore unitarity) cannot decrease the von Neumann entropy of the radiation demonstrating the correctness of the basic approach in~\cite{Mathur:2009hf}.}
%
\begin{figure}[htbp]
\begin{center}
\subfigure[]{
\includegraphics[scale=.5]{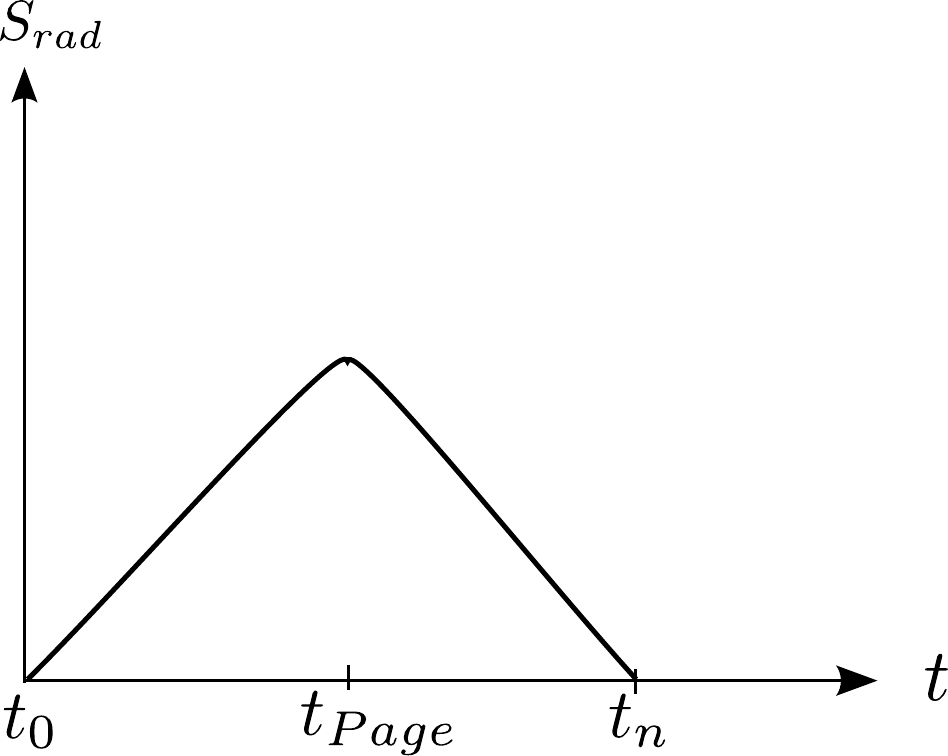}} \hspace{1in}
\subfigure[]{ 
\includegraphics[scale=.5]{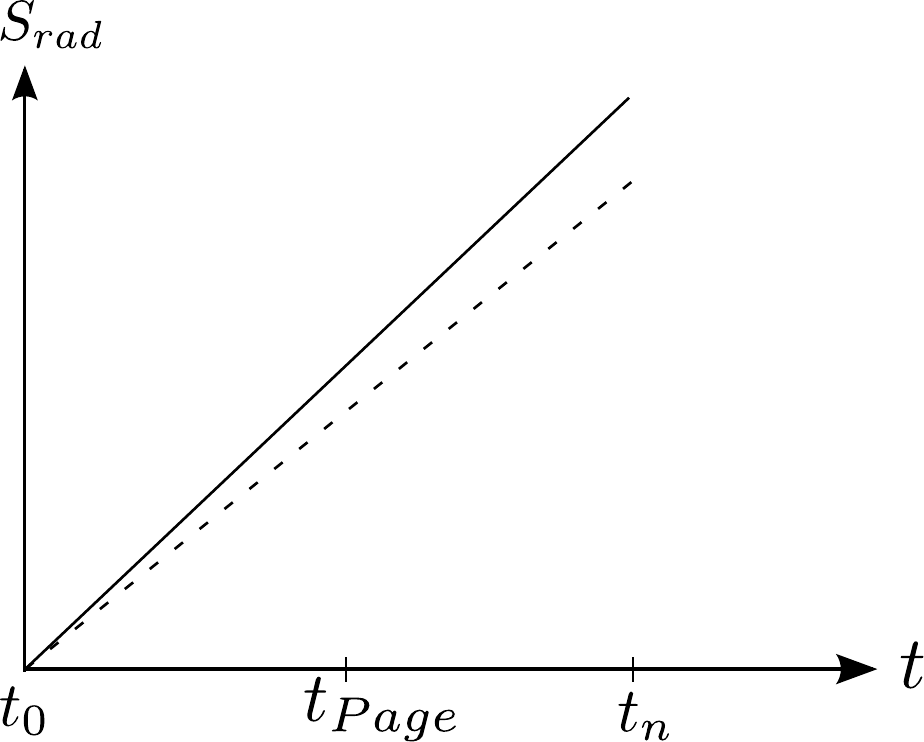} } 
\caption{
Entanglement entropy of radiation from  (a) a normal body and (b) a traditional black hole with information-free horizon. In normal burning bodies, assuming them to be in a typical state to begin with, the entropy initially goes up and then goes down while for traditional black holes that evaporate via Hawking-pair creation the entropy monotonically increases. Allowing small correction to the leading order process (solid line) decreases the slope (dashed line) but the entropy curve keeps rising. (For the specific curves shown, one should interpret the time coordinate on the horizontal axis as being chosen such that quanta are emitted at a constant rate per unit \emph{coordinate} time.)
}
\label{evaporation}
\end{center}
\end{figure}
In other words purity of the final state requires order one corrections (i.e. not suppressed in $M^{-1}$) to the Unruh vacuum. Thus, unitarity demands the black hole to have non-trivial horizon-scale structure in accord with the fuzzball program.

While this argument shows that small corrections to the leading order Hawking process cannot restore {\em purity} of the final state,  we would like to emphasize that modifying the process to ensure purity of the final state does \emph{not imply unitarity}. In other words, purity of the final state is a necessary but not a sufficient condition for unitary evaporation. 
One of the goals of this paper is to make this more precise and see its implications for the information paradox. In addition to purity of the final state, unitarity requires linearity, norm preservation and invertibility of evolution. With these conditions and the reasonable assumption that the dimension of the Hilbert space does not drastically change during the evaporation process, we show that information has to come out at every step during the evaporation process for unitarity to be preserved. 

At this point, it is important to note the distinction between information {\em decoding} and information {\em release}.  In ordinary evaporating systems there are correlations between successively emitted quanta that encode information of the structure of the surface they have been emitted from. While this information can only be read off after the radiation composes more than half of the total system the answer will depend on what has been emitted previously. In that sense information must have been carried away by the emitted quanta.
Using qubit models we will show that for black hole evaporation to be unitary {\em every quantum coming out of the black hole, before and after Page time, has to carry information of the black hole state} in a sense made more precise in Section~\ref{Qubit}.
Thus the state at the horizon\footnote{We do not attempt go give a precise definition here; we think the concept of a state at the \emph{horizon} as a classical location is ill-defined, and disappears in a full theory of quantum gravity.} cannot be in the Unruh vacuum even before Page time. This result supports the fuzzball conjecture.
\end{itemize}
Horizon-scale structure immediately raises the question we next turn to of what an infalling observer experiences.

\begin{itemize}
\item {\bf Infall}

BHC asserts that an infalling observer experiences nothing unusual while passing through the horizon, but how can this be consistent if unitarity requires the horizon to contain information? 
Since the question of  unitarity is addressed by outside observers whose experiences are conjectured to be complementary to infalling observers' experiences, Mathur's argument does not directly lead to a sharp contradiction with BHC. Almheiri--Marolf--Polchinski--Sully (AMPS)~\cite{Almheiri:2012rt}, however, recently showed that after the Page time one can make a modified version of Mathur's argument entirely in the infalling observer's frame and therefore conclude that the postulates of BHC are inconsistent. AMPS then propose that an infalling observer entering a black hole after the Page time hits a ``firewall'' before reaching the horizon.\footnote{Relatedly, Ref.~\cite{Braunstein:2009my}, using a different argument, posits that an infalling observer hits an "energetic curtain" at the horizon of a black hole; however it is suggested that this phenomenon may only arise once the black hole is planck-sized.}\footnote{There has been some discussion on whether the firewall is before, at or after the horizon. 
Further ``infall" is sometimes used to refer to falling towards the horizon and sometime through the horizon. In this paper we advocate that  spacetime should end outside the horizon in a fuzz of stringy modes and furthermore by infall we mean not just the approach to the fuzz but an {\em effective} description for $E \gg T_H$ quanta involving passing through the horizon.}

Let us now turn to the fate of infalling observers in the fuzzball proposal.
For the fuzzballs constructed so far in supergravity spacetime ends before the horizon and an incoming observer would either bounce off the bottom of the fuzzball or go splat depending on its details. The constructed solutions are single microstate geometries. 
Black holes, on the other hand, represent ensembles of microstates and so, to address the fate of an infalling observer, one must discuss the infall question for fuzzballs that are {\em typical} in Page's sense.
Since black hole evaporation originates from semi-classical evaporation of the infalling vacuum and typical Hawking quanta have  {Killing} energy $E \sim T_H$, it is clear that according to the fuzzball proposal such quanta cannot have a free infall into black holes. The question is whether one can  say something about the fate of quanta with $E \gg T_H$ without an explicit construction of typical fuzzballs. Recently, Mathur introduced \emph{fuzzball complementarity}~\cite{Mathur:2010kx}: while quanta with $E \sim T_H$ do not have free infall independent of the details of the fuzzball, more energetic quanta $E \gg T_H$ fall in freely, oblivious to the details of the fuzzball. This proposal, further developed in~\cite{Mathur:2011wg, Mathur:2012jk, Mathur:2012zp, Mathur:2012dx}, explains in what sense black holes are coarse-grained 
descriptions of 
fuzzballs. We review fuzzball complementarity in Section~\ref{Approximate Complementarity}.

While there seems to be growing consensus  that there are order one corrections to the semi-classical physics at the horizon scale, opinions differ on what the implications are for the infall problem. AMPS~\cite{Almheiri:2012rt} and Susskind~\cite{Susskind:2012uw, Susskind:2012rm} argue that after the Page time no modes can have free infall. Giddings argues for ``non-local" and ``non-violent" effects which give a free infall for all modes~\cite{Giddings:2011ks, Giddings:2012bm}. While Bousso affirms the basic correctness of the firewall argument, he suggests the possibility of large corrections to the semi-classical description motivated by the Horowitz--Maldacena final state proposal~\cite{Horowitz:2003he}. See also~\cite{Ori:2012jx, Banks:2012nn,Hossenfelder:2012mr}.  To the best of our knowledge, apart from fuzzballs, nowhere else is the energy scale dependence regarding the infall problem discussed. One of our main points is to emphasize the importance of this separation of scales.

\end{itemize}

The plan of this paper is as follows. In Section \ref{Qubit}, using qubit models, we present the main result of this paper that information is carried out with each emitted quantum.  There is \emph{no difference in the process and rate of information release before and after Page time}. This is used to argue that if unitarity is to be preserved, the configuration resulting from infalling matter can never be the traditional black hole geometry with Unruh vacuum at the horizon. A less technical discussion of the results of this section is provided in  Subsection \ref{LessonsLearnt}. In Section \ref{Approximate Complementarity} we work under the assumption that the true microstates of quantum gravity are fuzzballs and review the possibility of fuzzball complementarity where high-energy quanta ($E \gg T_H$) have a universal infall into fuzzballs but quanta of typical energy ($E \sim T_H$) do not have any such universal description. Thus it is made clear in what sense coarse-graining of typical fuzzball 
results in a black hole geometry. Finally, in Section \ref{Conclusion} we conclude with a summary of our results. 

\section{When and how does information {\em of the original state} come out?} \label{Qubit}

The purpose of this section is to study the evaporation of quantum mechanical qubit systems and the conditions required for a unitary evolution. We will then rephrase the Hawking process in this language  and render more precise the questions of why unitarity breaks down and what the necessary conditions are to restore it. 

\begin{figure}[htbp]
\begin{center}
\subfigure[]{
\includegraphics[scale=.55]{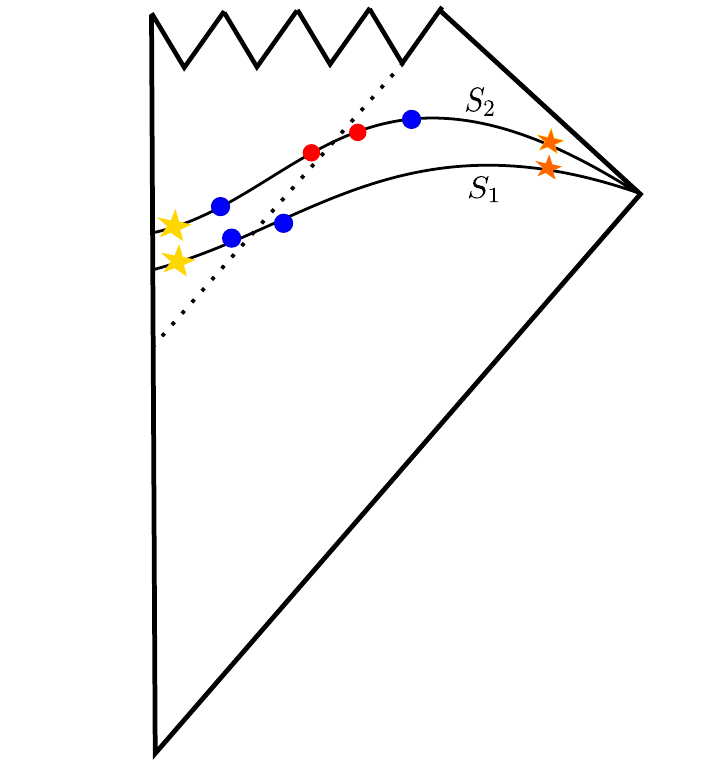}} \hspace{1in}
\subfigure[]{
\includegraphics[scale=.30]{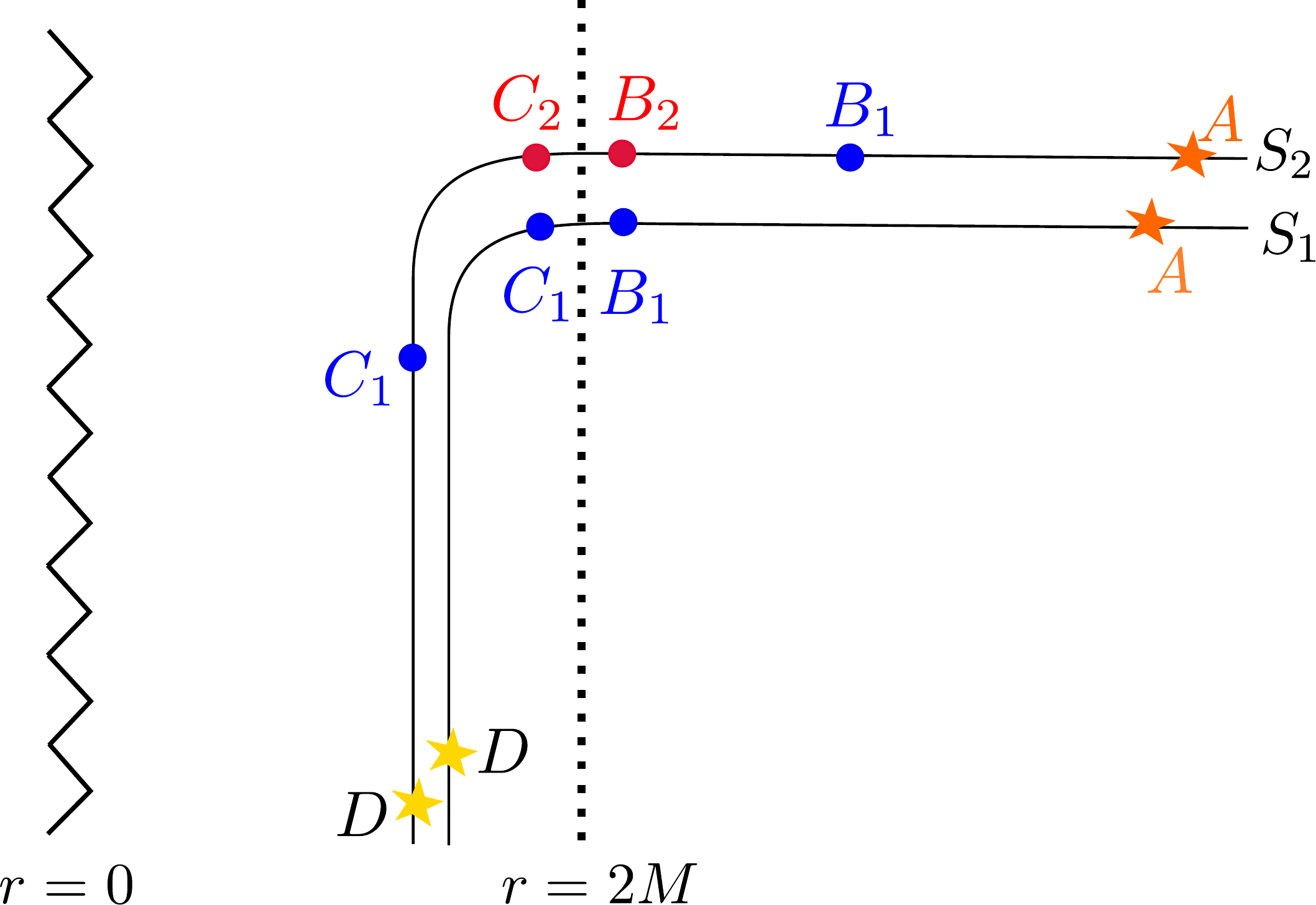}}
\caption{(a) Two subsequent ``nice slices'' that smoothly intersect initial infalling matter (yellow), newly created Hawking-pairs (blue and red) at the horizon and early Hawking radiation (orange). (b) Hawking-pair creation at the horizon (dashed line) on two subsequent nice slices sufficiently far away from the singularity (zigzag line) in order to avoid large curvatures. On slice $S_1$ the outgoing member of the pair created at the horizon is labelled ``$B_1$" and its ingoing partner is labelled ``$C_1$". On slice $S_2$ the pair $B_1C_1$ has moved away from the horizon in opposite directions and a new pair ``$B_2$" and ``$C_2$" is created. When not needed, we omit the subscripts which indicate  the slice on which they were created.
}
\label{niceSlices}
\end{center}
\end{figure}

To this end it is useful---although our analysis does not depend on it---to adopt the notation used for ``nice slices''. This foliation of the black hole spacetime 
avoids the strong curvature regions and yet cuts through the initial matter, horizon, and outgoing Hawking radiation in a smooth way~\cite{Lowe:1995ac}. Two such  slices, $S_1$ and $S_2$,  are depicted in the Penrose diagram in Figure \ref{niceSlices}a and in a schematic form in Figure \ref{niceSlices}b.
\footnote{The evaporation time scale of a black hole of mass $M$ is of order $\sim M^3$ which is parametrically larger than the time gap between successive quanta coming out which is of order $\sim M$.
So for the purpose of a few emissions we can use the nice slices in Figure \ref{niceSlices}a.} The initial matter, labelled ``$D$", that formed the black hole is in the deep interior while any  earlier emitted radiation, labelled ``$A$'', has moved far away from its place of creation at the horizon.  The 
evolution of a slice results in a stretching of the connector region straddling the horizon and this 
creates the Hawking-pair. 
For details of this process see \cite{Mathur:2008wi,Mathur:2009hf}. We use the letters $B$ and $C$ to denote  outside and inside Hawking quanta respectively. 

The pair created at the horizon is like an EPR pair.  While the infalling member of the pair might be ``caught'' by the initial matter, it is well known that the EPR pair cannot be used to communicate information. In similar vein, the information of the initial matter $D$ cannot be carried away by the outgoing member of the Hawking-pair. We give a more technical account of this phenomenon 
when rephrasing the Hawking process in the language of qubit models.

We use the notation $ABCD$ for the general study of quantum mechanical evaporation via qubit models and the conditions of unitarity without referring to a horizon. 
To make contact with Hawking evaporation we introduce ficticious ``pairs'' $BC$ that allow us to deform the unitary evaporation process into the Hawking process and phrase, in the qubit framework, the exact statement for the break-down of unitarity.  We argue that for evaporation to be unitary, $B$ has to carry away information of $D$ at every step of the evaporation.  For typical states (defined as states where each small subsystem is almost maximally entangled with the rest) at no point can the $BC$ system be in a pure state. For atypical states, while the $BC$ system may be in a pure state, it cannot be in a predetermined state independent of $D$. 
In the context of black holes this means that the state at the horizon can never be the Unruh vacuum state.

We now state the conditions required for unitary evaporation which our analysis is based on.
Regardless of the details of evaporation, there are at least four necessary conditions for unitarity of black hole evaporation:
\begin{enumerate}
\item Purity: Pure states evolve into pure state.	
\item Linearity: The map between initial and final states is linear.
\item Preservation of norm: Evolution of states preserves norm.
\item Invertibility: The map of black hole initial state to the radiation is invertible.
\end{enumerate} 
While condition 1 has been the focus of many recent works, the other conditions have largely not been discussed to the best of our knowledge. The main goal of this paper is to discuss the implications of the latter three,  especially of condition 4. However, before doing that we would like to summarize the recent discussions about condition 1.

\subsection{The role of purity}

Condition 1 was emphasized by Page in \cite{Page:1993up} where he argued that if the black hole is in a typical pure state and {\em one assumes unitary evaporation} as for any normal body then the entropy of the radiation at first increases and then decreases to zero again (depicted in Figure \ref{evaporation}a), so that the final state is indeed pure. This can be roughly understood as follows. In a typical state each small subsystem is almost maximally entangled with the remaining (larger) subsystem \cite{Page:1993df}. If we assume the state of the entire system to be pure then measuring the entropy of the radiation is like tracing over the remaining black hole system. As long as the radiation is the smaller part of the system its entropy increases. However, after half the system has evaporated, the black hole is the smaller system and the entropy of the radiation starts decreasing until it becomes zero again.

In a series of recent papers  \cite{Mathur:2009hf,Mathur:2010kx,Mathur:2011wg,Mathur:2011uj,Mathur:2012np} Mathur argued that a Hawking-{\em like} evaporation process, \emph{cannot} bend the entropy curve to go down at any point in time (see Figure \ref{evaporation}b).
The argument can be summarized as follows. Let $A$ be the previously emitted radiation, $B$ a newly created outgoing quantum, and $C$ its ingoing partner as shown in Figure \ref{niceSlices}. The statement of strong subadditivity \cite{Lieb:1973cp} for the system $ABC$ {with { independent} subsystems $A$,$B$ and $C$} is
\be
S_{AB} + S_{BC} \ge  S_B + S_{ABC}\,.  \label{subadditivity0}
\ee
If the evaporation happens by Hawking-pair production the state at the horizon, $BC$,  is the Unruh vacuum. In other words  $B$ and $C$ are maximally entangled: $S_{BC} = 0$. With this, $S_{ABC}=S_A$ and the above relation becomes
\be
S_{AB}  \ge  S_B + S_A\,.  \label{subadditivity1}
\ee
On the other hand the statement of subadditivity \cite{Araki:1970ba} is
\be
S_{AB} \le S_A + S_B\,. \label{subadditivity2}
\ee
Putting \bref{subadditivity1} and \bref{subadditivity2} together yields
\be
S_{AB} = S_A + S_B\,. \label{subadditivity3}
\ee
The systems $A$ and $B$ are thus not correlated and this implies $S_{AB} \ge S_A$ and the entropy of radiation outside never decreases. Actually Mathur's claim is stronger as he shows that small corrections  to the leading order Hawking process (those which vanish when we take $M \to \infty$), also do no turn the entropy curve around. This result is illustrated in Figure \ref{evaporation}b. In fact, Mathur~\cite{Mathur:2009hf} only explicitly considers one type of correction, which even when made arbitrarily large cannot turn the entropy curve around~\cite{Giddings:2012dh}. The quantitative part of Mathur's argument is therefore unconvincing. Fortunately, one of the authors generalized the result for all corrections, and explicitly considered a family of models that smoothly deforms the Hawking process into unitary evolution~\cite{Avery:2011nb}.\footnote{The specific unitary model that is deformed to was introduced in~\cite{Giddings:2011ks}, and further discussed in~\cite{Giddings:2012bm}.}

In \cite{Almheiri:2012rt} AMPS have interpreted the above result, in the context of black hole complementarity, to argue for a phenomenon which they dub the ``firewall". Their argument consists of two parts:
\begin{enumerate}
\item {\em To ensure purity of the final state of radiation} it is necessary  that $S_{AB} < S_A$ after Page time. Thus one has $S_{BC} \ne 0$ no later than the Page time.
\item An observer falling in after Page time will encounter a blue shifted $B$ in her own frame and thus will burn at the horizon.
\end{enumerate}
While an infalling observer always encounters the system $B$ in a blue shifted state, if the $BC$ system is the Unruh vacuum state, the experience of the infalling observer is drama (and trauma) free. However, $S_{BC} \ne 0$ precludes this possibility and thus AMPS argue that it is not possible for an infalling observer to experience free infall through the horizon after Page time.

This concludes our summary of the recent discussion of the role of condition 1 in the recent debates on the information loss paradox. In the following we argue that the non-vanishing of $S_{BC}$ after Page time is a \emph{necessary but not sufficient} condition to preserve unitarity. Specifically, to ensure the conditions 1--4 mentioned earlier $S_{BC}$ \emph{cannot be zero during any part of the evaporation process} for typical states. Furthermore, when the state is not typical, $S_{BC}$ may be vanishing but the $BC$ system cannot be in a predetermined state independent of the black hole state $D$. Thus specifically the $BC$ system cannot be in the Unruh vacuum state at any point during the evaporation. 

In the discussion below we assume that the dimensionality of the Hilbert space is preserved during evaporation and that the Bekenstein--Hawking entropy is a measure of the number of pre-collapse configurations~\cite{PhysRevLett.49.1683}. Possible concerns regarding these assumptions are discussed in Subsection~\ref{LessonsLearnt}.

\subsection{A very simple ``moving-bit" model for evaporation} \label{VerySimpleModel}

Following~\cite{Mathur:2011wg, Mathur:2011uj, Mathur:2010kx, Mathur:2009hf, Giddings:2011ks, Giddings:2012dh, Giddings:2012bm, Mathur:2012zp, Czech:2011wy} we model the black hole evaporation process through the evolution of qubits.
We start with a simple intuitive model before discussing the qubit formalism more generally in Subsection~\ref{QubitModel}.
Consider simple unitary evolution of a system that moves, qubit by qubit, from location $x$ to location $y$. We start with a system of $n$ qubits at position $x$,
\be
\ket{\psi_{0}} = \ket{D_n^x} \otimes \dots \otimes \ket{D_1^x} = \bigotimes_{j=n}^{1} \ket{D_j^x}\,, \label{initialState}
\ee
where each $\ket{D_j}$ is a qubit  and the superscript $x$ denotes the location of a bit. Each $\ket{D_j}$ factor thus actually represents 2 qubits worth of information: 0 or 1, and $x$ or $y$. One could imagine this as the Hilbert space of $n$ spin-$\frac{1}{2}$ distinguishable particles on a 2-site lattice with all particles starting on one site.
For illustrative purposes, we have taken a direct product of qubits; the evolution of arbitrary superpositions, including typical states, is determined by the evolution of this basis of states. 
At the first step of the evolution the first bit gets moved from location $x$ to location $y$, $\ket{D_1^x} \stackrel{\U_1}{\too} \ket{D_1^y}$. We can write $\U_1$ explicitly as
\begin{equation}
\U_1 = I\otimes ( \ket{1_1^y}\bra{1_1^x} + \ket{0_1^y}\bra{0_1^x} +  \ket{1_1^x}\bra{1_1^y} + \ket{0_1^x}\bra{0_1^y})\,, \label{eqn:MovingOperator}
\end{equation}
where $I$ is the identity which acts on all the other qubits. In later expressions, the identity should be assumed to act on any subspaces not shown.
Subsequent evolution operators $\U_2$ through $\U_n$ have the above action on qubits $2$ through $n$, respectively.
By the $i$th step, the above initial state has evolved to
\be
\ket{\psi_{i}}=\prod_{j=i}^{1} \U_j \ket{\psi_0} =\bigotimes_{j=n}^{i+1} \ket{D_j^x}\otimes \bigotimes_{k=i}^{1} \ket{D_k^y}\, \qquad i \ge 1.\label{intermediateState}
\ee
The final state after $n$ steps,
\be
\ket{\psi_n}=\prod_{j=n}^{1} \U_j \ket{\psi_0} = \bigotimes_{j=n}^{1} \ket{D_j^y}\,,\label{finalState}
\ee
is thus just a relocation of the original state. 

In this way we can view the  ``moving bit" model as a simple evaporation process.  Having demonstrated how the basis vectors in~\eqref{initialState} evolve; the evolution of superpositions of the states of that form follow from the above. 

This kind of evaporation looks qualitatively different from black hole evaporation, where it seems that at each step two extra qubits are created. 
However since the infalling member of the Hawking-pair has negative energy it is supposed to somehow cancel with a quantum of the original matter.  In order to put unitary models like the ``moving bit'' model and potentially non-unitary pair creation models on the same footing, one of the authors introduced a general model space in which at each time step two extra qubits are created, just like one might model Hawking-pair creation, but one also must define two internal black hole qubits to be auxiliary~\cite{Avery:2011nb}. By tracing out the auxiliary degrees of freedom one gets a description of the physical, fixed-dimensional Hilbert space. Thus, the auxiliary degrees of freedom are simply a convenient way of parametrizing potentially non-unitary evolution, such as occurs in models of Hawking-pair creation. By the end of the evolution, all of the physical degrees 
of freedom are radiation, and therefore by definition the remaining black hole degrees of freedom are all auxiliary. More details can be found in Subsection \ref{QubitModel}. 

We would now like to illustrate how one can reinterpret the above model as a pair creation process using the ideas from~\cite{Avery:2011nb}.
Since the model we give below is simply a rewriting of the above model, it is manifestly unitary. In the ``moving bit'' model we are  considering, two auxiliary qubits are introduced at each step  as follows. The first step of the  ``evaporation'' process of \eqref{initialState} is  given by 
\begin{equation}
\mathcal{I}_1 \ket{\psi_0} = \bigotimes_{j=n}^{2} \ket{D_j^x} \otimes \ket{d_1^x} \otimes \ket{c_1^x} \otimes \ket{B_1^x}\,,
\end{equation}
where $\ket{B_1^x}$ and $\ket{c_1^x}$ are the created pair. The two lower-case qubits,  $\ket{d_1^x}$ and $\ket{c_1^x}$, are the auxiliary qubits.
To match our ``moving bit'' model we have $B_1 = D_1$. Since we are introducing new degrees of freedom, $\mathcal{I}_1$ is not unitary; however, conservation of probability  demands that it be isometric, that is to say it must satisfy $\mathcal{I}_1^\dagger\mathcal{I}_1 = I$. This constraint along with $B_1 = D_1$ ensures that the states $\ket{d_1^x}$ and $\ket{c_1^x}$ must be independent of the initial state $\ket{\psi_0}$. (If this point is not obvious, it is elaborated on in Subsection~\ref{BleachBleachBleach}.) When this is true, we say that the auxiliary qubits are in fiducial form, ``bleached'', or ``zeroed''.  Let us comment that the fiducial form can be any predetermined state, so long as it is totally independent of the initial conditions. It should be clear, then, that this step is a trivial rewriting of the state $\ket{\psi_0}$.

This is followed by the unitary transformation $\U_1^{'}$ relocating the qubit $\ket{B_1^x}$ to location $y$ via $\ket{B_1^x} \stackrel{\U_1^{'}}{\too} \ket{B_1^y}$. Hence, the state after emission of one qubit is
\begin{equation}
\ket{\psi_1} = \U_1^{'} \mathcal{I}_1 \ket{\psi_0}= \bigotimes_{j=n}^{2}  \ket{D_j^x}\otimes \ket{d_1^x} \otimes \ket{c_1^x} \otimes \ket{B_1^y}\,.
\end{equation}
Similarly, the state at the $i$th step is
\be
\ket{\psi_i}
=\bigotimes_{j=n}^{i+1} \ket{D_j^x} \otimes \bigotimes_{k=i}^{1}\Big(\ket{d_k^x} \otimes \ket{c_k^x} \Big)\otimes \bigotimes_{m=1}^{i} \ket{B_m^y}\,,
\ee
and the final state is
\be
\ket{\psi_n} =\bigotimes_{j=n}^{1}\Big(\ket{d_j^x} \otimes \ket{c_j^x} \Big) \otimes \bigotimes_{m=1}^{n} \ket{B_m^y}.\label{finalStateEvap}
\ee
Since the auxiliary qubits must all be in fiducial form given by some state $\ket{\phi}$,\footnote{Note that the fiducial state could take step dependent values, however the algorithm assigning the values must be independent of the initial state.} we can write them as
\be
\ket{d_i^x} \otimes \ket{c_i^x} = \ket{\phi} \otimes \ket{\phi} \quad \forall ~ i
\ee
so the final step may be written
\begin{equation}\label{eq:final-state-2}
\ket{\psi_n} =\bigotimes_{j=1}^{2n}\ket{\phi} \otimes \bigotimes_{m=1}^{n} \ket{B_m^y}.
\end{equation}

After tracing out the auxiliary degrees of freedom in \eqref{eq:final-state-2} and recalling that for our simple model $\ket{B^y_i} = \ket{D^y_i}$ we see that the final state is just the relocated initial state \eqref{initialState}. In this simple model of evaporation using auxiliary qubits one can easily see that information leaves the system in \emph{every step} of the evaporation process.

Let us note that for the specific evolution shown above the entanglement $S_{B_ic_i}$ is zero at every step; however, the evolution is for a very special basis of states of the form~\eqref{initialState}. A generic state will be a superposition of states of that form. Since the $c_i$ are always in fiducial form, they are direct-producted into the state. Therefore $S_{B_ic_i} = S_{B_i}$, but since we are just moving qubits this is equivalent to computing $S_{D_i}$ in the initial state. We can then appeal to Page~\cite{Page:1993df} to argue that generically for large $n$,  $D_i$ is maximally entangled with the rest of the degrees of freedom. Thus, in this model, for typical initial states  $S_{B_ic_i}$ is nonzero \emph{at every step}. For atypical initial states, even though $S_{B_ic_i}$ may be zero at some steps, the $B_i c_i$ system is not in a predetermined state 
which is independent of the initial state.

The important technical lesson here is that unitarity requires the newly emerging quanta (that leave the system) to carry information of the old quanta (that are part of the interior of the system) and avoiding quantum cloning means that the old quantum has to be bleached. Unitarity also requires the auxiliary quanta $c_i$ to be bleached. We give a more detailed account of this in Subsections \ref{QubitModel} and \ref{BleachBleachBleach}.

\subsection{Qubit models of evaporation} \label{QubitModel}

The simple model in the previous section serves as an example of unitary evolution that demonstrates for typical states $S_{BC}\neq 0$ at every step of the evolution. For atypical states while $S_{BC}$ may be zero for some steps, the $BC$ system is not in a predetermined  state independent of the initial state. Thus in all cases the $BC$ system cannot be in a predetermined state which is independent of the state of the evaporating system during any step of the evaporation. We buttress the above model with a second example. The model below explores what happens if the $BC$ system is in the Hawking-state (and thus has $S_{BC}=0$) until Page time, at which point one allows for more general evolution. 

\subsubsection*{General quibt model structure}
We start by reviewing the general structure of these qubit models. For more details and more models the reader is referred to \cite{Avery:2011nb} (see \cite{Mathur:2011wg, Mathur:2011uj, Mathur:2010kx, Mathur:2009hf, Giddings:2011ks, Giddings:2012dh, Giddings:2012bm, Mathur:2012zp, Czech:2011wy} for related models). The initial matter is modeled as a set of $n$ ``matter qubits'':
\be
\rho_0 = \ket{\psi_0} \bra{\psi_0} \qquad \ket{\psi_0} \in \text{span} \{\ket{\hat{q}_1\cdots\hat{q}_n}\}
\ee
where each $\hat q$ is a qubit. After a sequence of intermediate steps the final state consists entirely of radiation, modeled as a (possibly mixed) density matrix, $\rho_f$, acting on $n$ radiation qubits. Throughout the evolution we keep the dimension of the physical Hilbert space fixed. We discuss this point further in Subsection~\ref{LessonsLearnt}.

Following~\cite{Giddings:2011ks}, we use hats to distinguish the internal black hole qubits from the external radiation qubits. The hatted qubits represent all degrees of freedom that are inaccessible outside the black hole; unlike \cite{Mathur:2009hf}, we do not distinguish between degrees of freedom from the initial matter, from gravitational interactions, or any that arise during the evaporation process. One could model general (not necessarily unitary) discrete time evolution on the $n$-qubit Hilbert space in two ways~\cite{Avery:2011nb, nielsen}: as a sequence of mappings\footnote{Physically reasonable requirements suggest that the mappings should be trace-preserving quantum operations~\cite{nielsen}. Thus, the two descriptions correspond to the operator--sum representation and the  environmental-coupling representation of quantum operations.} on  $n$-qubit density matrices, or alternatively as unitary 
transformations on pure states in an enlarged Hilbert space. In typical applications, one 
imagines that the 
additional degrees of freedom in the enlarged Hilbert space represent environmental degrees of freedom; however, since we are considering potentially fundamental non-unitarity, the additional degrees of freedom are just ``auxiliary'' degrees of freedom. Motivated by the latter and typical discussions of the pair creation process, we model the dynamics as a sequence of isometric mappings in an enlarging Hilbert space. The extra degrees of freedom are auxiliary variables that provide a convenient language to discuss a more general evolution.
While we keep the dimension of the \emph{physical} Hilbert space fixed, at each step of emission we introduce two auxiliary qubits. Thus on the $i$th step of the evolution, the total state is described by a $n+2i$ qubit state $\ket{\psi_i}$ with $2i$ hatted auxiliary qubits and $i$ radiation qubits. At each step $i$, the physical state may be recovered by tracing over the auxiliary space
\be
\rho_i = tr_{aux}\left(\ket{\psi_i}\bra{\psi_i}\right)\,.
\ee
We model the evolution in two steps: a creation step effected by
isometric operators $\mathcal{I}_i$;\footnote{These were called $C_i$ in~\cite{Avery:2011nb}.} and an internal evolution step effected by
unitary $\hat{U}_i$ acting on the hatted qubits and unitary $U_i$ acting on the
unhatted radiation qubits. Basis vectors at each step look like
\begin{multline}
\text{span}\left\{\ket{\hat{q}_1\hat{q}_2\cdots\hat{q}_{n+i}}\ket{q_iq_{i-1}\cdots q_1}\right\}
 \xrightarrow{\mathcal{I}_i} 
\text{span}\left\{\ket{\hat{q}_1\hat{q}_2\cdots\hat{q}_{n+i}\hat{q}_{n+i+1}}\ket{q_{i+1}q_iq_{i-1}\cdots q_1}\right\}\\
\xrightarrow{\hat{U}_i\otimes U_i}
\text{span}\left\{\ket{\hat{q}_1\hat{q}_2\cdots\hat{q}_{n+i}\hat{q}_{n+i+1}}\ket{q_{i+1}q_iq_{i-1}\cdots q_1}\right\}\,.
\end{multline}
The qubits are arranged in the above order to match with a crude notion of locality expected on the nice slices~\cite{Mathur:2009hf,Giddings:2011ks}.

For the majority of the discussion, we focus on the $\mathcal{I}_i$ and are
content to set $\hat{U}_i = U_i = I$. What properties should the $\mathcal{I}_i$
satisfy?  Conservation of probability and the linearity of quantum mechanics suggest that the $\mathcal{I}_i$ should preserve the norm and be linear, which
means that we should require
\begin{equation}\label{eq:CdgC}
\mathcal{I}_i^\dagger \mathcal{I}_i = I\,,
\end{equation}
where there is no sum on $i$.  Note that $\mathcal{I}_i$ is an isometric, but 
non-unitary mapping. We also assume that, apart from the newly created pair,  the $\mathcal{I}_i$ act only on the
hatted black hole qubits and not on the unhatted radiation qubits
which are far away from the pair creation site.
We can write the $\mathcal{I}_i$ in the following form
\begin{equation}
  \mathcal{I}_i = \ket{\varphi_1}\otimes \hat{P}_1+\ket{\varphi_2}\otimes \hat{P}_2+\ket{\varphi_3}\otimes \hat{P}_3 + \ket{\varphi_4}\otimes \hat{P}_4\,,
\end{equation}
where $\ket{\varphi_j}$ are an orthonormal basis for the created pair
qubits, and the $\hat{P}$'s are linear operators which act on the
previous hatted qubits (with implicit $i$ dependence). {Since there are no operators acting on the unhatted radiation qubits, the evolution of qubits outside the black hole is local.}
Following~\cite{Mathur:2011wg, Giddings:2011ks}, we use the basis
\begin{equation}\begin{aligned}
\ket{\varphi^i_1} &= \frac{1}{\sqrt{2}}\big(\ket{\hat{0}_{n+i+1}}\ket{0_{i+1}} 
                           + \ket{\hat{1}_{n+i+1}}\ket{1_{i+1}}\big)\,,\\
\ket{\varphi^i_2} &= \frac{1}{\sqrt{2}}\big(\ket{\hat{0}_{n+i+1}}\ket{0_{i+1}} 
                           - \ket{\hat{1}_{n+i+1}}\ket{1_{i+1}}\big)\,,\\
\ket{\varphi^i_3} &= \ket{\hat{0}_{n+i+1}}\ket{1_{i+1}}\,, \\
\ket{\varphi^i_4} &= \ket{\hat{1}_{n+i+1}}\ket{0_{i+1}}\,,
\end{aligned}\end{equation}
for the newly created pair. The constraint in Equation~\eqref{eq:CdgC}
implies the following condition on the $\hat{P}$s:
\begin{equation}\label{eq:completeness}
\mathcal{I}_i^\dagger \mathcal{I}_i = \hat{P}_1^\dagger\hat{P}_1 + \hat{P}_2^\dagger\hat{P}_2 
+ \hat{P}_3^\dagger\hat{P}_3 + \hat{P}_4^\dagger\hat{P}_4 = \hat{I}\,.
\end{equation}
Note that this defines the $\hat{P}$s as a set of generalized
measurement operators acting on the black hole Hilbert space. 
A fully specified model, then, entails
\begin{enumerate}
\item A set of $\hat{P}$s at each step $i$ that satisfy the completeness
  relation~\eqref{eq:completeness}.
\item The unitary operators $\hat{U}_i$ and $U_i$ for each $i$.
\item A clear delineation of the auxiliary subspace at each step $i$.
\end{enumerate}
 This gives a very general model space that makes it easy to compare and contrast
different models of evolution. 
We investigate the conditions for unitarity of the evolution in more detail in Subsection~\ref{BleachBleachBleach}, but let us now give the basic picture.
If one wants to acquire the fixed-dimensional Hilbert space description then one must trace out the auxiliary degrees of freedom at each step. Thus, in unitary models the auxiliary degrees of freedom must be direct-producted into the physical state. Since in a unitary model the physical state is by definition a unitary transformation from the previous physical state and since there can be no entanglement with auxiliary degrees of freedom for all previous physical states, the auxiliary degrees of freedom must be put into a fiducial state. That is to say, in unitary models the auxiliary degrees of freedom must be bleached of all information. Regardless of the details of the model, at the end of $n$-steps, when the black hole is supposed to have evaporated away, all of the hatted qubits are auxiliary and should be traced over to recover the physical state of the radiation.

\subsubsection*{A pure but non-unitary evaporation}
With this brief overview on qubit models we are ready to discuss (non-unitary) models.
We consider a model which has $S_{BC}=0$ for the first half of the evolution (until Page time) and $S_{BC}\neq0$ in the second half. In this model the entropy of radiation before Page time goes up while after Page time it goes down again. We consider an initial state of $n$ qubits,
\begin{equation}
\ket{\hat{q}_1\cdots\hat{q}_n}.
\end{equation}
For simplicity, let us take $n$ even. For the first $n/2$ timesteps let the pair creation be governed by exactly the Hawking-pair creation process
\begin{equation}
\mathcal{I}_i = \ket{\varphi_1^i}\otimes \hat{I}\quad \text{for} \quad i \leq \frac{n}{2}\,.
\label{eq:Cbefore}
\end{equation}
In each subsequent step we want to emit the information of all the earlier ingoing qubits from the black hole; this is achieved by mapping the information of an earlier ingoing qubit to a newly created outgoing qubit while bleaching the former via the operator
\bea
\mathcal{I}_i &=& \f{1}{\sqrt 2} ( \ket{ \varphi^i_1} + \ket{ \varphi^i_2} ) \otimes \ket{\hat 0} \bra{\hat 0 }_{\f{n}{2}+i}  + \ket{\varphi^i_3}   \otimes \ket{\hat 0} \bra{\hat 1 }_{\f{n}{2}+i} \nn
&=& \ket{{\hat{0}}{0}}_{\text{pair} }\otimes \ket{{\hat{0}}}\bra{{\hat{0}}}_{\frac{n}{2}+i}
       +\ket{{\hat{0}}{1}}_\text{pair}\otimes \ket{{\hat{0}}}\bra{{\hat{1}}}_{\frac{n}{2}+i} \quad \text{for} \quad i > \frac{n}{2}\,.
\label{eq:Cafter}
\eea
The $\hat P_i$ act on the $(\frac{n}{2}+i)$th black hole qubit (which is the infalling member of one of the Hawking-pair states created during the first $\f{n}{2}$ states). The information of this qubit is mapped to the $i$th outgoing qubit of the newly created pairs while the $(\frac{n}{2}+i)$th black hole qubit has to be set to a fiducial value which we choose to be zero. In addition the new infalling qubit is also bleached to zero.   This part of the evolution is equivalent to a model without auxiliary qubits that just emits (i.e. removes the hat from) the $(\frac{n}{2}+i)$th qubit; that is to say, it is equivalent to the ``moving bit'' model introduced in Section~\ref{VerySimpleModel}.
Let us work through a 4 qubit example of the evolution to illustrate the dynamics: 
\begin{equation}
\begin{aligned}
\ket{{\hat{q}_1 \hat{q}_2 \hat{q_3} \hat{q}_4}}\qquad \quad &\\
\stackrel{i=1}{\longrightarrow}\quad & \frac{1}{\sqrt{2}}\ket{{\hat{q}_1 \hat{q}_2 \hat{q_3} \hat{q}_4}}(\ket{\hat{0}0}+\ket{\hat{1}1})\\
\stackrel{i=2}{\longrightarrow}\quad&\frac{1}{2}\ket{{\hat{q}_1 \hat{q}_2 \hat{q_3} \hat{q}_4}}
      (\ket{{\hat{0}}\hat{0}00}+\ket{{\hat{0}}\hat{1}10}+\ket{{\hat{1}}\hat{0}01}+\ket{{{1}}\hat{1}11})\\
\stackrel{i=3}{\longrightarrow}\quad&\frac{1}{2}\ket{{\hat{q}_1 \hat{q}_2 \hat{q_3} \hat{q}_4}\hat{{0}}}
      (\ket{{\hat{0}}{\hat{0}}{0}00}+\ket{{\hat{1}}{\hat{0}}{0}10}+\ket{{\hat{0}}{\hat{0}}{1}01}+\ket{{\hat{1}}{\hat{0}}{1}11})\\
\stackrel{i=4}{\longrightarrow}\quad&\frac{1}{2}\ket{{\hat{q}_1 \hat{q}_2 \hat{q_3} \hat{q}_4}{\hat{0}\hat{0}\hat{0}\hat{0}}}
      (\ket{{00}00}+\ket{{10}10}+\ket{{01}01}+\ket{{11}11}).
\end{aligned}
\label{eq:4qubit}
\end{equation}
In the first two steps evaporation happens by the Hawking mechanism and in the subsequent two steps the previously created infalling member of the Hawking-pair is spat out. It is easy to verify that the von Neumann entropy of the radiation behaves like Page's prediction in Figure \ref{evaporation}a. However, it is also easy to see  that the final state, obtained by tracing out the hatted black hole qubits, is independent of the initial state. This evolution is not invertible and thus not unitary.

Note that this model has a pure final state. It also allows for the state of the created pair to be the Hawking-pair state for the first half of the evolution thus allowing for free infall of observers through the horizon until Page time.  This clearly demonstrates that purity of the final state of radiation does not necessarily ensure unitarity of the evaporation process.

\subsection{Conditions for unitarity} \label{BleachBleachBleach}

Before going into conditions for unitary evolution of the physical degrees of freedom, let us first discuss isometries in generality for a moment. Consider an isometric operator $\mathcal{I}$ that maps states from one Hilbert space $\mathcal{H}_1$ to another (at least as big) Hilbert space $\mathcal{H}_2$. Let us introduce an orthonormal basis of states for $\mathcal{H}_1$, $\{\ket{\hat{w_i}}\}$, then the most general isometry may be written as
\begin{equation}
\mathcal{I} = \sum_{i}\ket{v_i}\bra{\hat{w}_i},
\end{equation}
where $\ket{v_i}\in \mathcal{H}_2$. At this point, we have not said anything about the states $\ket{v_i}\in\mathcal{H}_2$; however, the isometry condition is
\begin{equation}
\hat{I}= \mathcal{I}^\dagger\mathcal{I} = \sum_{i,j}\ket{\hat{w}_i}\braket{v_i|v_j}\bra{\hat{w}_j}.
\end{equation}
Writing $\hat{I} = \sum_i\ket{\hat{w}_i}\bra{\hat{w}_i}$, we conclude that the $\ket{v_i}$ must be orthonormal. Let the range of $\mathcal{I}$ be the subspace $\mathcal{V}\subset \mathcal{H}_2$ where $\dim\mathcal{V}=\dim\mathcal{H}_1 \leq \dim\mathcal{H}_2$.  We can  decompose $\mathcal{H}_2$ as $\mathcal{H}_2 = \mathcal{V}\oplus \mathcal{V}^\perp$.  Moreover, when $\dim \mathcal{H}_1$ divides $\dim\mathcal{H}_2$, as is true for qubits, we may decompose  as $\mathcal{H}_2 = \mathcal{V}'\otimes \mathcal{B}$, where $\mathcal{V}'$ is isomorphic to $\mathcal{V}$. The kets $\ket{v_i}$ then decompose as
\begin{equation}
\ket{v_i} = \ket{v'_i}\otimes\ket{b},
\end{equation}
where all $\ket{v_i}$ have the same $\ket{b}\in\mathcal{B}$. This is all to say that the isometry $\mathcal{I}: \mathcal{H}_1\to\mathcal{V}'\otimes\mathcal{B}$ must act on $\ket{\hat{w}}\in\mathcal{H}_1$ as
\begin{equation}\label{eq:gen-isom}
\mathcal{I}:\ket{\hat{w}}\mapsto \big(U\ket{\hat{w}}\big)\otimes \ket{b},
\end{equation}
where $U$ is a unitary transformation and $\ket{b}$ does not depend on the state $\ket{\hat{w}}$. This follows from the no-cloning theorem. Thus every isometry can be given by specifying the bleaching subspace $\mathcal{B}$, its fiducial state $\ket{b}$, and the unitary transformation $U$. Therefore, every isometry that gives nonunitary evolution necessarily bleaches a bleaching subspace $\mathcal{B}$ with dimension $\dim\mathcal{B} =\dim\mathcal{H}_2/\dim\mathcal{H}_1$---an isometry giving unitary evolution has an empty bleaching space B.

Returning to qubit models of black hole evaporation, recall that to describe the physical $n$-qubit space, we need to trace out the ``auxiliary'' subspace. On step $i$, let the  $2i$-qubit auxiliary subspace be $\mathcal{A}_i$. Suppose that the evolution of the \emph{physical} degrees of freedom up to step $i$ have been unitary. This means that tracing $\mathcal{A}_i$ out of $\ket{\psi_i}$ gives a pure state:
\begin{equation}
\text{tr}_{\mathcal{A}_i}\left( \ket{\psi_i}\bra{\psi_i}\right) = \ket{\Psi_i}\bra{\Psi_i},
\end{equation}
where $\ket{\Psi_i}$ describes the $n$-qubits worth of physical degrees of freedom. Moreover, unitarity means that $\ket{\Psi_i}$ is a unitary transformation of the initial $n$-qubit state $\ket{\psi_0}$:
\begin{equation}
\ket{\Psi_i} = U\ket{\psi_0}.
\end{equation}
The cumulative isometry from the initial state $\ket{\psi_0}$ to the state $\ket{\psi_i}$ given by $\mathcal{I}_{i}\mathcal{I}_{i-1}\cdots \mathcal{I}_{1}$ maps a $n$-qubit space into a $(n+2i)$-qubit space, thus the bleaching subspace $\mathcal{B}$ of the cumulative isometry must coincide with the auxiliary subspace $\mathcal{A}_i$. If this were not true, then tracing out $\mathcal{A}_i$ would not give a pure state for all possible initial states $\ket{\psi_0}$. This follows from the general form of an isometry in Equation~\eqref{eq:gen-isom} and the no-cloning theorem.

Now, consider the evolution from $\ket{\psi_i}$ to $\ket{\psi_{i+1}}$. The isometry $\mathcal{I}_{i+1}$ maps the  $(n+2i)$-qubit space into a $(n+2i+2)$-qubit space and thus has a 2-qubit bleaching space $\mathcal{B}$. By assumption, we know that $\mathcal{I}_{i+1}$ acts as the identity on the $i$ radiation qubits outside the black hole. If $\mathcal{I}_{i+1}$ is going to generate unitary evolution of the physical degrees of freedom, then tracing out $\mathcal{A}_{i+1}$ needs to give a pure state $\ket{\Psi_{i+1}}$, which is also a unitary transformation of $\ket{\Psi_i}$. One can then see that the 2-qubit bleaching space of $\mathcal{I}_{i+1}$ must coincide with the 2-qubit space added to $\mathcal{A}_i$ to form $\mathcal{A}_{i+1}$.

By contrast, evolution for which the von Neumann entropy of the created pair \mbox{($S_{BC}=0$)} vanishes for all possible ``input states'', $\ket{\psi_i}$, is necessarily non-unitary. 
If the created pair is not entangled with the rest of the system for all $\ket{\psi_i}$, then the creation operator must be of the form
\begin{equation}
\mathcal{I} = \ket{\psi_\text{pair}}\otimes \hat{P}\,.
\end{equation}
Thus, the pair state space corresponds to the bleaching space $\mathcal{B}$ and $\hat{P}$ must be unitary in order to satisfy~\eqref{eq:CdgC} so that probability is conserved. On the other hand, we know that at each time step two new qubits must become part of the auxiliary space. If the physical evolution is going to be unitary, then that auxiliary space must be part of the bleaching space, i.e. projected into a fiducial form.  As discussed in~\cite{Avery:2011nb}, there are two slightly different ways to do this: either both new auxiliary qubits can be chosen from the old hatted qubits, or the new black hole qubit in the pair state can be put in fiducial form along with one old hatted qubit.  Let us note that the second possibility already implies that the pair is not in the Hawking-state $\ket{\varphi^i_1}$.  Regardless, unitarity of the physical evolution demands that $\mathcal{I}$ projects at least one qubit's worth subspace of the old black hole qubits into fiducial form; however, it is not 
simultaneously possible for $\hat{P}$ 
to be unitary and to project out a two-dimensional subspace. Hence the Hawking-pair production always leads to a non-unitary evolution.

If one does not take the above conditions for unitarity at intermediate steps seriously and only wishes to discuss unitarity in terms of the initial and final states, then one may be tempted to circumvent the above argument in the following way. One might suggest that the evolution does not bleach qubits on some early steps in the evolution, and then makes up for it at the end of the evolution by bleaching more degrees of freedom; however, this too is not possible. At each step, one only has enough freedom to bleach at most 2 qubits worth of degrees of freedom, and there are only $n$ steps to bleach $2n$ black hole qubits. The freedom to bleach 2 qubits comes precisely from the 4-dimensional pair space. This is exactly why the model in Section~\ref{QubitModel} failed to be unitary. It should be clear that while our discussion uses qubits, the arguments apply more generally. 

Alternatively, one might suggest that demanding that the pair has vanishing entanglement for \emph{all} input states is too restrictive. This is especially true if one subscribes to the argument advanced in~\cite{Giddings:2009gj,Susskind:2012rm} that young black holes are ``special.'' How special do young black holes need to be?
If the 4-dimensional bleaching space is all old black hole qubits (the first possibility mentioned in the previous paragraph), then the $\ket{v_i}$ in Equation~\eqref{eq:gen-isom} must span the remaining $2^{n+i}$ dimensional space \emph{including the 4 pair states}. Thus, the subspace which produces the Hawking state has codimension $4$ with the total space of input states assuming unitarity \emph{at each step} of the evolution. A model of this kind is proposed in~\cite{Giddings:2011ks}, along with the suggestion that the black hole starts in a specific state. We discuss this possibility further in Subsection~\ref{LessonsLearnt}.

\subsection{Where is the information?} \label{LocalizedInformation}

We can more precisely quantify what happens to the information by using the standard quantum information technique of entangling degrees of freedom with a fictitious reference system~\cite{nielsen}. We follow the approach of Giddings and Shi~\cite{Giddings:2012dh}, which refers to earlier work in this context~\cite{Hayden:2007cs}. In order to quantify what then happens in the evaporation process to all possible initial states, we can introduce a ``tracking state'' that is maximally entangled between the black hole initial state and the reference system:
\begin{equation}
\ket{T} = \frac{1}{2^{\frac{n}{2}}}\sum_{i=1}^{2^n}\ket{\hat{w}_i}\ket{\tilde{r}_i}\,,
\label{eq:trackstate}
\end{equation}
where $\{\ket{\hat{w}_i}\}$ is an orthonormal basis for the $n$-qubit initial state, and $\{\ket{\tilde{r}_i}\}$ is an orthonormal basis for the $n$-qubit reference system. We then evolve the black hole degrees of freedom using the evolution under study, while acting on the reference system by the identity. This allows us to keep track of the fate of each initial state---the location of the initial entanglement with the reference system tells us the fate of the information.

As discussed in~\cite{Giddings:2012dh}, using the state $\ket{T}$, the decrease of the von Neumann entropy of the black hole degrees of freedom quantifies how much information has \emph{left} the black hole; whereas, the von Neumann entropy of the external radiation quantifies how much information is now in the radiation alone. In general, these two quantities need not be equal, since information may be stored in the correlations between the two subsystems. The amount of extra correlation is given by the mutual information between the two subsystems, which~\cite{Giddings:2012dh} conjectures to be zero for black hole evaporation. When the mutual information vanishes, Giddings and Shi term the evolution ``subsystem transfer''~\cite{Giddings:2012dh}. Perhaps the simplest example of subsystem transfer would be the moving bit model in Subsection~\ref{VerySimpleModel}.

In fact, as we now show, in our qubit framework one finds that any unitary evolution necessarily has vanishing mutual information between the black hole and the radiation with the tracking state $\ket{T}$. Let the space of black hole qubits at step $i$ be $D_i$, and the space of radiation qubits be $A_i$. Recall from the discussion in the previous section that unitarity demands that at every step of the evolution two qubits of $D_i$ be bleached of information, while only one hatted qubit is created. Thus, at step $i$ the effective dimension of $D_i$ is $2^{n-i}$-dimensional, and therefore
\begin{equation}\label{eq:SQ}
S(D_i) \leq (n-i)\log 2.
\end{equation}
On the other hand subadditivity implies that
\begin{equation}\label{eq:subadd}
S(D_i) + S(A_i) \geq S(D_i A_i) = n\log 2\,,
\end{equation}
where the last step follows from the form of the tracking state and the reference degrees of freedom not mixing with the rest of the system. Together~\eqref{eq:SQ} and~\eqref{eq:subadd} imply
\begin{equation}
S(A_i) \geq i\log 2\,.
\end{equation}
Since the space of radiation qubits is $2^i$-dimensional, it follows that
\begin{equation}
S(A_i) = i\log 2\,,\qquad S(D_i) = (n-i)\log 2
\end{equation}
for all $i$. Thus, we find that unitary evolution implies that information leaves the black hole 
at \emph{every step} via ``subsystem transfer''. Let us note that the requirement may be relaxed slightly if one supposes that the dimension of the physical Hilbert space increases in the evaporation process, so that there are more than $n$ steps of evolution for the information to come out; however, one still finds that information comes out from the beginning of the evolution if the extra steps are evenly interspersed.  Even if one frontloads the extra steps, according to arguments of \cite{PhysRevLett.49.1683} one still only anticipates 30\% extra steps and information still starts coming out well before Page time. We discuss our assumptions about the size of the Hilbert space more in Subsection~\ref{LessonsLearnt}.

One may be confused by the above statement since Page's result~\cite{Page:1993df} is often colloquially stated as showing that information starts leaving the black hole only after the black hole has evaporated halfway. In fact, Page's result shows that one may start \emph{reconstructing} the original black hole state only after Page time, i.e. after the black hole has evaporated halfway. This is actually a different statement. Page's result is further weakened by the observations in~\cite{Hayden:2007cs} that after Page time new information entering the black hole may be recovered in the black hole scrambling time, and black holes have subsequently been conjectured to be the fastest physically achievable scramblers in nature~\cite{Sekino:2008he}.

Using the above tracking state, we may also quantify how much information has been lost in the case of non-unitarity evolution. In particular, the von Neumann entropy of the auxiliary subsystem $\mathcal{A}_i$ measures the entanglement between the reference system and the auxiliary degrees of freedom. In other words, for the tracking state $\ket{T}$, $S(\mathcal{A}_i)$ gives the amount of information lost. 
Let us illustrate this with an example.
Let the initial black hole state with basis $\{\ket{\hat{w}_j}\} \in D_0$ be a 2-qubit system that is maximally entangled with a reference system system with basis $\{\ket{\tilde{r}_j}\}\in R_0$ in a tracking state \eqref{eq:trackstate} that is initially of the form
\be
\ket{T_0}=\frac{1}{2} \Big( \ket{\hat{0}\hat{0}}\ket{\tilde{0}\tilde{0}} + \ket{\hat{0}\hat{1}}\ket{\tilde{0}\tilde{1}}  +  \ket{\hat{1}\hat{0}} \ket{\tilde{1}\tilde{0}}+  \ket{\hat{1}\hat{1}}\ket{\tilde{1}\tilde{1}}\Big)\,.
\label{eq:tack}
\ee 
Let the evolution be governed by an isometry that mimics the moving bit model of Subsection \ref{VerySimpleModel}:\footnote{Note that \eqref{eq:trackisometry} is the same evolution operator as \eqref{eq:Cafter}.}
\be
\mathcal{I}_i = 
 \Big(\ket{\hat{0}\hat{0}}\ket{0}\bra{\hat{0}}_{(3-i)} + \ket{\hat{0}\hat{0}}\ket{1}\bra{\hat{1}}_{(3-i)}\Big)_{D_{i-1}} \otimes I_{A_{i-1}}\otimes I_{R_{i-1}}
\quad \text{for} \quad i = 1,2\,
\label{eq:trackisometry}
\ee
where the subscript $(3-i)$ on the RHS indicates the action on the $(3-i)$th hatted qubit of the black hole system $D_{i-1}$ at the $i$th step and the action on the reference system $R_{i-1}$ and the outside radiation $A_{i-1}$ is the identity. 
The first step of the evolution yields
\be
\ket{T_1}=\mathcal{I}_1 \ket{T_0}=\ket{\hat{0}\hat{0}} \otimes  \frac{1}{2} \Big(\ket{\hat{0}0}\ket{\tilde{0}\tilde{0}}+\ket{\hat{0}1}\ket{\tilde{0}\tilde{1}}+\ket{\hat{1}0}\ket{\tilde{1}\tilde{0}}+\ket{\hat{1}1}\ket{\tilde{1}\tilde{1}}\Big)\,.
\ee
By comparison with \eqref{eq:gen-isom} we can see that the direct producted state comes from the bleaching space $\mathcal{B}_1$.
The unhatted qubits are the radiation qubits in $A_1$; which of the hatted qubits are part of the physical space of black hole qubits $D_1$ after tracing out the auxiliary space $\mathcal{A}_1$ is part of the model. Here we are interested in 
how much information we lose if we trace over the remaining hatted qubit of the initial state and one of the qubits of the bleaching space.
We can easily verify that the thus obtained density matrix \mbox{$\rho_1={\rm Tr}_{\mathcal{A}_1} \ket{T_1}\bra{T_1}$} has ${\rm Tr}\rho_1^2<1$ and is thus mixed. How much information we lost by tracing out part of the initial state is encoded in $S(\mathcal{A}_1)$. The density matrix in the auxiliary space is
\be
\rho_{1}^{aux}=\ket{\hat{0}}\bra{\hat{0}} \otimes \frac{1}{2}\Big(\ket{\hat{0}}\bra{\hat{0}}+\ket{\hat{1}}\bra{\hat{1}}\Big)\,,
\label{eq:densitymatrix}
\ee
with von Neumann entropy 
\be
S({\mathcal{A}_1})=-{\rm Tr}(\rho_{1}^{aux}{\rm log} \rho_{1}^{aux})=\log 2\,.
\ee
The auxiliary space thus contains the maximal information of the traced out qubit that has been part of the initial black hole state.
After the second evolution step we identify the auxiliary space $\mathcal{A}_2$ with the bleaching space $\mathcal{B}_2$ and the final state of the radiation is given by the density matrix 
\be
\rho_2^{rad}={\rm Tr}_{\{R_2,\mathcal{A}_2\}} (\mathcal{I}_2 \rho_1 \mathcal{I}^\dagger_2)=\frac{1}{2}\Big(\ket{00}\bra{00}+\ket{01}\bra{01}\Big)\,.
\ee
While in this simple example we have chosen the initial state to be a 2-qubit system the extension to $n$-qubit systems is straightforward. This example shows that if, at any step $i$ of the evolution, we trace out a qubit that is maximally entangled with the remaining system we lose the maximal amount of information of $\log 2$ on that qubit encoded in the von Neumann entropy $S(\mathcal{A}_i)$ of the auxiliary space at that step. While this simple example results in the maximum amount of information loss on a qubit, more general examples should have $0<S(\mathcal{A}_i)<\log 2$ for each physical qubit that has erroneously been traced out.
That unitary evolution should have no information loss and therefore $S(\mathcal{A}_i)=0$ for the tracking state, immediately implies that the auxiliary space must coincide with the bleaching space in accord with the discussion in the previous subsection. 

Let us emphasize that the interpretations of the entanglement entropies discussed in this section only hold when considering the state $\ket{T}$.

\subsection{Discussion of the results} \label{LessonsLearnt}

Since most of the discussion in this section was technical we now recap the main result, mention the assumptions we have made and analyze similarities and differences with ideas of some other works.

\subsubsection*{The result}

Black hole evaporation, or at least its traditional description, is unlike other more familiar evaporation processes. Consider two boxes A and B. A quantum being moved from one box to another can be modeled by the following Hamiltonian:
\be
\mathcal H_{int} = \sum_{ij} (C_{ij} a_i^\dagger b_j +C_{ij}^*  a_i  b_j^\dagger) \label{normal}\,,
\ee
where we see that creation of a particle in box A is accompanied by annihilation of one in box B and vice versa. This can be used as a simple model of evaporation. The evaporating body's internal excitations annihilate to emit the outgoing particles. However in the pair-creation model of Hawking radiation, when a bit of radiation is created outside the black hole, another bit is {\em created} inside. In our two box system this would be:
\be
\mathcal H_{int} = \sum_{ij} (C_{ij} a_i^\dagger b_j^\dagger +C_{ij}^*  a_i  b_j). \label{pairProduction}
\ee
While this might seem odd at first sight, this kind of interaction occurs in situations other than black holes as well. For example this reminds us of  Schwinger pair production \cite{PhysRev.82.664} and the Schiff--Snyder--Weinberg effect \cite{PhysRev.57.315}. In the latter a strong background electric field is turned on in a region separating box A and box B. This causes particles to be produced in box A and anti-particles to be produced in box B.  From conservation of energy we know that the interaction in Equation~\eqref{pairProduction} cannot be the whole story. Indeed, one can view the background potential in room B as coming from lots of particles. Loosely, we can say these particles then leak out to room A. Then \bref{pairProduction} looks just like a fancy way of writing \bref{normal}. Of course, we do not mean to imply that particle number is always conserved. Only that \emph{something} gets annihilated, and \emph{something} gets created. For the black hole, we know that the initial mass that 
formed the black hole plays the role of the field in Schwinger pair creation. 

In the above examples, the Hilbert space is infinite-dimensional Fock space. Furthermore, throughout the evolution, there is no sense in which the dimension of the Hilbert space changes; however, if we go from a field theory description to a description in terms of single particle states in multiparticle quantum mechanics, then the dimension of the Hilbert space apparently increases throughout the dynamics in~\eqref{pairProduction} and stays fixed throughout the dynamics in~\eqref{normal}. What is the significance of this difference? Let us think about a high-energy collision where a lot of particles are produced, the relevant interaction is schematically $C_{ijklm\cdots}(a_{i}b_{j}c_{k}^\dagger d_{l}^\dagger e_{m}^\dagger\cdots)$. Again, there seems to be a large increase of the dimension of the Hilbert space in a multiparticle quantum mechanical description; however, we know this must be illusory since the S-matrix is unitary. Indeed, for a given initial state, we expect the final state to be a given 
entangled multiparticle state. If we return to~\eqref{pairProduction}, the increase in the Hilbert space is an illusion since the produced pair are not independent of the quanta being created in room A and the one being annihilated in room B. For an application of such a process to describe fuzzball evaporation see~\cite{Chowdhury:2008bd}.

Let us note that there is another way of reducing Fock space to a finite-dimensional Hilbert space. One often can restrict one's attention to a finite-dimensional subspace by appealing to conservation of energy and applying appropriate cutoffs. So, we may think about the qubit models more abstractly, as modeling the dynamics on the constraint surface. Then, we assume the dynamics we discuss include \emph{all} the degrees of freedom: not just field theory modes, but also quantum gravitational degrees of freedom. The interpretation of our results may then be a bit less direct, however, we expect the particle interpretation outside the black hole to continue to hold. Translating from the above mentioned multiparticle quantum mechanics to this kind of description should correspond to tracing out auxiliary degrees of freedom to get a fixed-dimensional Hilbert space.

It turns out that any evaporation process which looks like~\eqref{normal} can be written in a way where it seems to occur by pair creation~\eqref{pairProduction} by introducing auxiliary quanta. However, unitarity puts strict conditions on what state the auxiliary quanta can be in. As an example, we studied a simple ``moving-bit" model in Subsection~\ref{VerySimpleModel}. At each step of evolution two extra auxiliary qubits are added to the system. While this model is trivial, it helps elucidate what conditions the auxiliary qubits need to satisfy for unitarity. We then studied more general qubit models and put our results on a firmer footing. Our results indicate that auxiliary qubits need to be put in a fiducial form at each step of the evolution. In addition since the outgoing bit needs to carry information of the system being burnt, it cannot be auxiliary. It turns out that the bits which have to be auxiliary are the analog of the Hawking infalling quanta and one of the original qubits. With this result 
we 
were 
able to show 
that in our qubit models the ``pair" cannot be in a  pure state for arbitrary initial states if unitarity is preserved. 

Translated back to the black hole language, this means that the state at the horizon cannot be in the Unruh vacuum, or any other predetermined pure state independent of the microstate of the black hole, \emph{at any point during the evaporation}, if unitarity is to be preserved. This result supports the fuzzball proposal which states that the microstates accounting for the Bekenstein--Hawking entropy are horizonless and singularity-free.  We comment on what this implies for the infall problem in Section~\ref{Approximate Complementarity}.

\subsubsection*{Assumptions we made}
We have worked with four main assumptions which we list here explicitly so as to help make the comparison of our results with other models of quantum black holes easier.
\begin{itemize}
\item {\em The dimension of physical Hilbert space is constant} \\
The first  assumption we made in our work is that the dimension of the physical Hilbert space does not change. While one would certainly expect this to be true for any unitary theory, and thus for quantum gravity in particular, one may still be concerned that the qubit model is not able to capture all the Hilbert space when the volume in which the radiation is being emitted is infinite. In particular the irreversibility of free streaming  Hawking radiation leads to about 30\% extra entropy in the radiation compared to the black hole \cite{PhysRevLett.49.1683}\footnote{We are thankful to Samir Mathur for pointing this out to us.} (at the same time it is also shown in \cite{PhysRevLett.49.1683} that if the black hole is allowed to evaporate in a reversible manner the entropy of the radiation equals that of the black hole). Thus one may wonder if there is a possible loophole in our argument for free streaming Hawking radiation. While it is indeed true that the qubit model is unable to capture free streaming 
processes and  one may think that some fraction of the  created pairs may be in a special state (although for true free streaming they would not be in a special state but a random state), we would expect such events to be interspersed and not to affect our general argument. Furthermore, this cannot get one all the way to Page time.

\item {\em The black hole initial state is not special} \\
Another assumption we made is that the black hole formed is itself not in a special state to begin with, otherwise the emitted radiation would also be in a special state. There have been some papers advocating that in a suitably fine-grained description the number of degrees of freedom available to ordinary matter is parametrically smaller than the entropy of the black hole \cite{Giddings:2009gj,Susskind:2012rm}. We find this claim puzzling for two reasons. In a large enough box radiation can have arbitrarily many fine-grained configurations. Adiabatically collapsing the box does not change the number of possible configurations and the final state of the collapse, the black hole, can thus not be in a predetermined state. This idea is supported by Zurek's calculation \cite{PhysRevLett.49.1683} that a black hole can be allowed to evaporate in a reversible manner.
One may object that a black hole is not ordinarily made in such a controlled fashion but rather by collapsing a shell. However, the fundamental properties of black holes should not depend on how they were formed. Thus, we focussed on black holes formed in generic states. Because of the adiabatic nature of collapse, such models necessarily describe scrambled states. 
Secondly, even though black holes formed by ``normal'' astrophysical collapse may only occupy a small part of the total initial state space, it seems strange to posit that these states coincide with the states that are treated specially by the \emph{dynamics}, especially in light of the fast scrambling conjecture.

\item {\em The evaporation process can be described within a local framework}\\
In our discussion of black hole evaporation via qubit models we did not allow for any form of non-local effects that might somehow map the information to the outside radiation. While one might argue that a theory of quantum gravity will be non-local to some extent, one would have to specify the degree of non-locality outside the horizon and explain how it alters (or why it does not alter) physics in other systems consistent with experiments. For example Giddings, in a series of papers \cite{Giddings:2012dh,Giddings:2012gc,Giddings:2013kcj}, argues that information can be recovered if we give up the notion of \emph{local} quantum field theory as an exact framework. It is not clear, however, how unitarity can be preserved this way without drastically changing the thermodynamics of the black hole and the evaporation time scale. The purpose of this work was to formulate the black hole evaporation process in a local framework of qubit models and deduce the consequences of unitarity restoration.


\item\textit{The definition of the interior spacetime is state-independent}\\
  In our discussion of infall, we implicitly assume that the modes $B$ and $C$ must be in one particular ``safe state'',
  $\ket{\varphi_1}$, for the infalling observer to ``see'' the Unruh vacuum. Moreover, we assume that the $C$ qubit in
  the discussion is predetermined. One may imagine trying to evade some of the arguments in AMPS and here by supposing
  that the interior/horizon spacetime is \emph{defined} in a state-dependent way. For example, we might say that we take
  the last emitted $B$ and look for the qubit with which it is maximally enatangled, and \emph{define} that to be
  ``$C$'', and the entangled state to be the ``safe state'' $\ket{\varphi_1}$. Before the Page time the $C$ one arrives
  at via this procedure is at least still a black hole degree of freedom, but after the Page time it gives a $C$ from the
  radiation and is therefore nonlocal. This state-dependence appears to be a kind of nonlinear quantum
  mechanics, which we do not consider here.\footnote{In fact, a far more serious problem with such models is that they lead to loss of unitarity~\cite{Chowdhury:2013mka}.}


\end{itemize}

\section{Fuzzballs and the possibility of ``fuzzball complementarity"} \label{Approximate Complementarity}

We now want to explore the issue of infall into black hole microstates. It seems that many hold the view that the fate of an infalling quantum is universal and thus, irrespective of the energy, it experiences free infall through the horizon. We start our discussion of the infall question by challenging this view. We briefly explain the fuzzball proposal and then speculate on the what a microscopic description of falling into a typical fuzzball would entail. We then review fuzzball complementarity and  how it bypasses the AMPS argument.

\subsection{Is infall universal?}\label{sec:mirror}

\begin{figure}[htbp]
\begin{center}
\subfigure[]{
\hspace{2.8cm}\includegraphics[scale=.4]{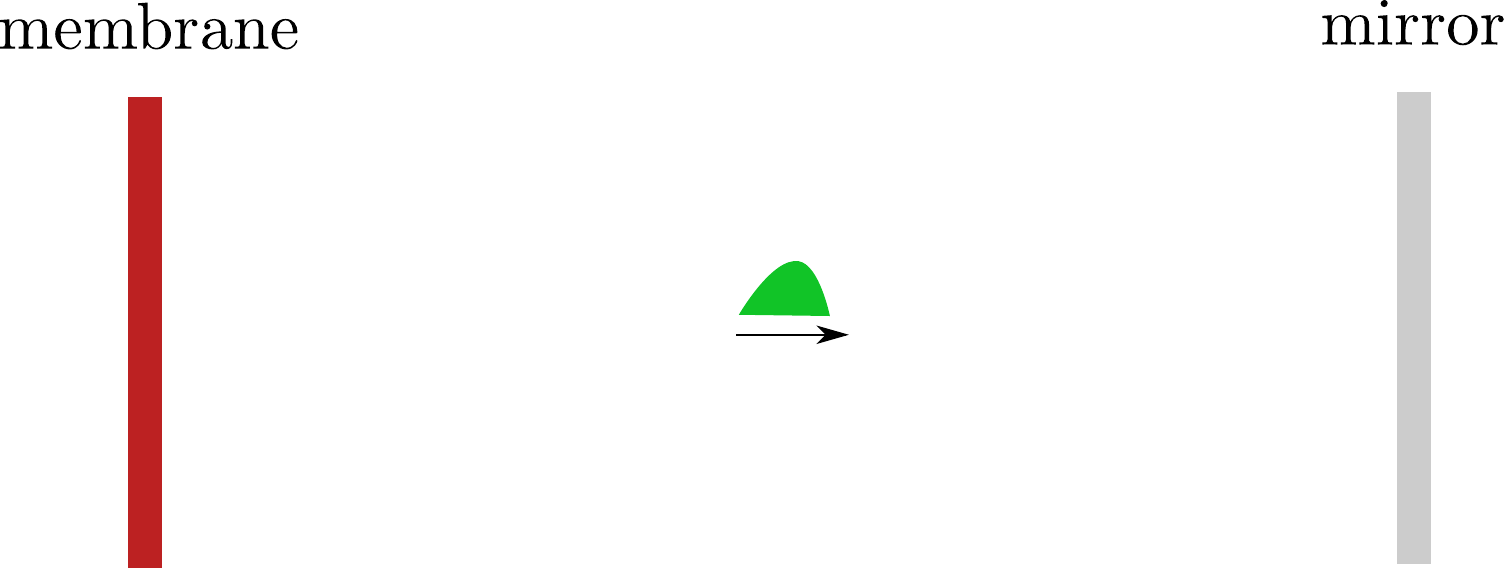}} \\
\subfigure[]{
\includegraphics[scale=.4]{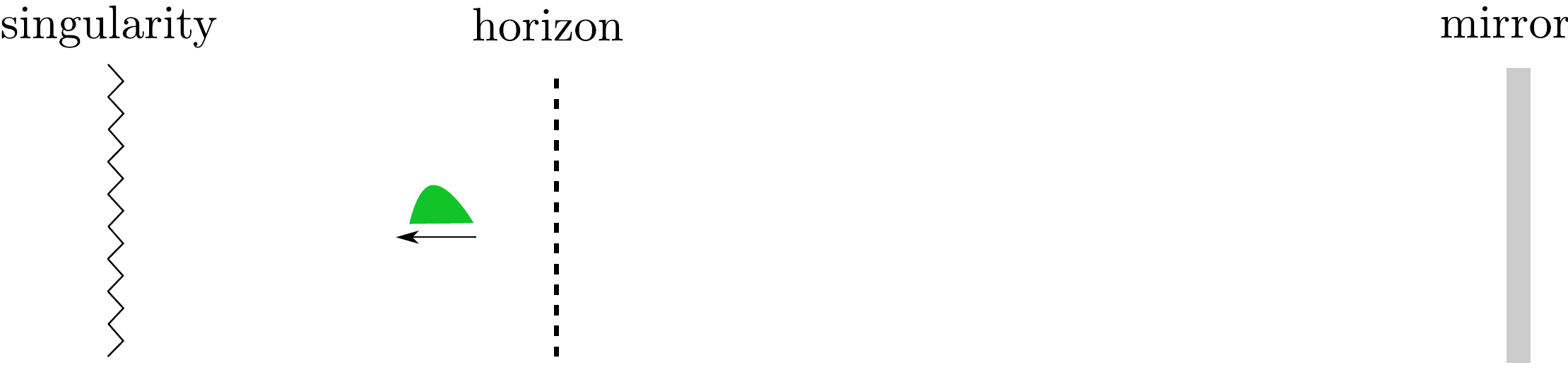}}
\caption{(a) A typical quantum comes out of the system. To preserve unitarity it must carry information of the state. According to black hole complementarity it originates from a hot membrane outside the event horizon. (b) According to views held by many if such a quantum is reflected back in it will have a free infall, independent of the details of the state. This seems odd because if there is a microscopic description of the emission process one would expect the microscopic dynamics to be reversible.}
\label{mirror}
\end{center}
\end{figure}

In the first part of this paper we argued that for unitarity to be preserved in the evaporation process 
every quantum coming out of the black hole has to carry information out.
Keeping this result in mind one can consider the following Gedankenexperiment shown in \mbox{Figure \ref{mirror}}. 
When a quantum comes out of the black hole, a mirror can be placed to throw it back in. While for unitary evaporation of a black hole the spectrum of outgoing radiation cannot be exactly thermal, we still expect information  to be mostly carried by the typical quanta which have Killing energy $E \sim T_H$. After being reflected from the mirror back towards the black hole it seems unreasonable to think that such a quantum will freely fall through the horizon since, when coming out it carried information of the state. Basically, if there is a microscopic description of the emission process, we expect the microscopic dynamics to be reversible. This reasoning, however, only applies to typical quanta and it may well be that the physics of infalling observers of Killing energy $E \gg T_H$ is very different. 
Such a separation in scales is manifest in Mathur's conjectured fuzzball complementarity which we discuss in detail in Section \ref{fuzzballcomplementarity}.

\subsection{Fuzzballs} \label{fuzzball}

In Section \ref{Qubit} we argued that purity of the final radiation state is not sufficient to guarantee a unitary evolution and showed that additional conditions make the argument for horizon-scale structure considerably stronger. We showed that at no point can there be a predetermined pure state (like Unruh vacuum) at the horizon. While Unruh and Hartle-Hawking vacuum have regular stress tensor at the future horizon, any other state is expected to have a divergent stress tensor and we thus need to look at the microstates of quantum gravity. To answer the question of what happens to an infalling observer one must therefore understand the nature of these microstates.

\subsubsection*{Information release} According to the fuzzball proposal the true microstates accounting for the Bekenstein--Hawking entropy have neither a singularity nor a horizon shielding it. A fuzzball is well approximated by a black hole asymptotically and, presumably, all the way to the potential barrier that surrounds the black hole. Beyond the barrier a fuzzball starts differing from a black hole at a location that depends on the particular fuzzball and ends in a, possibly non-geometric, complicated stringy ``fuzz'' somewhere between the potential barrier and the location of the would-be horizon.  
This proposal nicely resolves the information paradox as the radiation now emerges from the surface of these fuzzballs and thus carries information of them, much like the radiation emerging from the surface of a star or a piece of coal. Unitary evolution is thus built into fuzzballs by construction. For a review of the proposal see \cite{Mathur:2005zp,Bena:2004de,Skenderis:2008qn,Balasubramanian:2008da,Chowdhury:2010ct} and to see how radiation from fuzzballs carries information see \cite{Chowdhury:2007jx,Avery:2010hs}.

\subsubsection*{Infall experience}
Replacing the black hole solution by a fuzzball naturally raises the question: 
What is the experience of an infalling observer?
Infall into certain non-typical fuzzballs have been studied in \cite{Lunin:2001dt,Giusto:2004ip}. An observer falling into such a fuzzball bounces off at the bottom of the fuzzball behind which there is no spacetime. In \cite{Bena:2012zi} near-extremal supergravity fuzzballs have been constructed perbatively. They exhibit an erratic force on certain probes and since the experience of such an infalling probe into these fuzzballs will be violent one may interpret them as ``fuzzballs of fire''.

To make contact with the firewall argument we need to study infall into fuzzballs that are  { typical} in Page's sense.\footnote{It was pointed out in \cite{Susskind:2012rm} that the time for the full system to become generic is the recurrence time $\sim M e^{M^2}$. Note that this is parametrically larger than the evaporation time $\sim M^3$ so the system never becomes generic. However, an infalling observer experiences a small part of the system and thus by typical we mean scrambled. In \cite{Sekino:2008he} scrambling time is conjectured to be $\sim M \log M$. After scrambling time we expect to be able to approximate a small part of the system by a thermal density matrix for appropriate coarse-grained operators.} 
The mirror argument of Section \ref{sec:mirror} suggests that we should further split the infall question according to observers\footnote{A simple model for an observer is the Unruh-de Witt detector \cite{Unruh:1983ms}.} with Killing energy $E \sim T_H$ and those with $E \gg T_H$.
Since typical radiation quanta ($E \sim T_H$) do not come from the evolution of Unruh vacuum but from the surface of the fuzz, it is clear that at least infalling quanta of such energies can not have free infall. 
The fate of infalling quanta with Killing energy $E \gg T_H$ is not so obvious. 
In the remainder of this section we elaborate on a recent conjecture of Mathur where he claims that infall into {\em typical} fuzzballs 
can be approximated by infall into black holes for these high-energy modes. We elaborate on the sense in which coarse-graining of typical fuzzballs may give the black hole.

\subsection{Falling into typical fuzzballs} \label{TypicalFuzzballs}
\begin{figure}[htbp]
\begin{center}
\includegraphics[scale=.45]{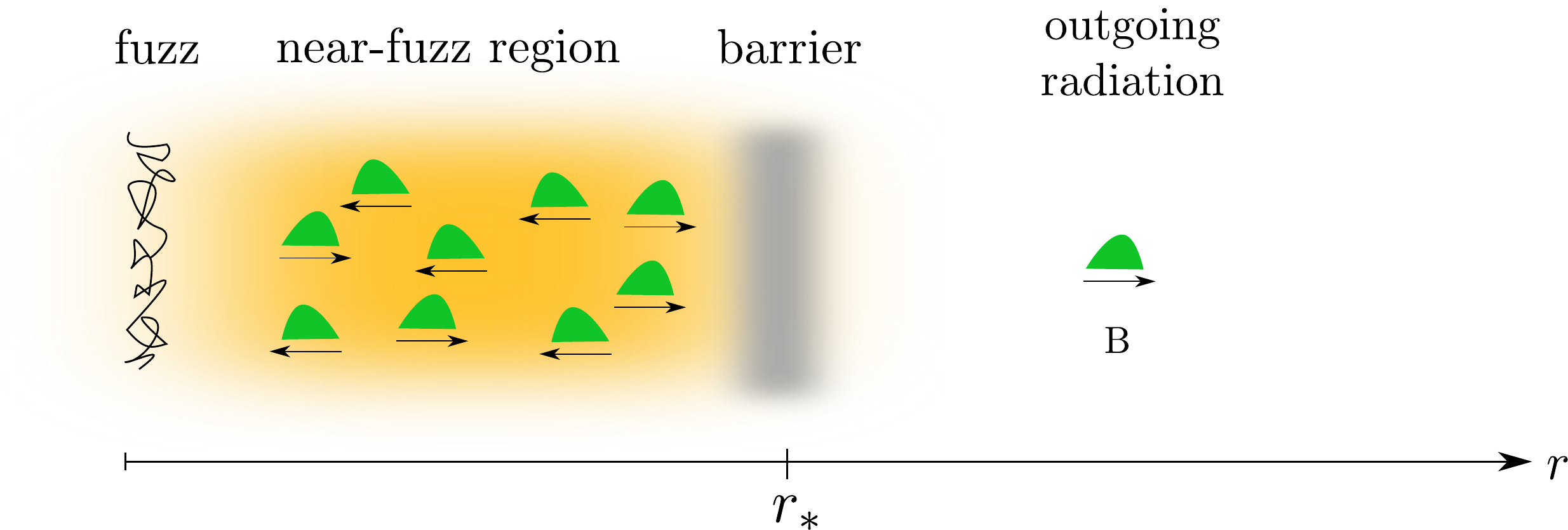}
\caption{To answer the infall question we have to discuss the interaction of the infalling observer with the full system: radiation quanta emitted from the fuzzball either cross the potential barrier at $r_*$ or get reflected back into the fuzz. An infalling observer thus encounters a single radiation quantaumoutside the barrier but a lot of quanta partially trapped between the barrier and the fuzz. Further they encounter the fuzz itself. We refer to this region as ``near-fuzz" instead of ``near-horizon" to emphasize the lack of a horizon in the fuzzball picture.}
\label{greybody}
\end{center}
\end{figure}

In this section we will speculate on the structure of typical fuzzballs and their similarities with and differences from black holes.
In the traditional black hole picture it is well known that the perfect thermal nature of the emitted radiation is ``spoilt" by gray-body factors. These arise from  the potential barrier outside the horizon through which the Hawking quanta have to tunnel to escape to infinity. While the exact form of this barrier is angular momentum dependent it peaks around a distance $\sim \mathcal{O}(M)$ outside the horizon. In the fine-grained version the horizon is not there and the fuzzball ends with a boundary, the fuzz, which is somewhere between the potential barrier and the location of the would-be horizon. The picture expected for a typical fuzzball is illustrated in \mbox{Figure \ref{greybody}} where we denote the location of the barrier by $r_*$. A  quantum that leaves the fuzz has a non-vanishing probability of getting reflected by the potential barrier back towards it. Thus after the scrambling time, we expect a gas of quanta filling the region between the fuzz and the barrier in dynamic 
equilibrium with the fuzz. However, if this system is embedded in flat space then any outgoing quantum which manages to escape to the other side of the potential barrier escapes to infinity.  
We refer to the region between the fuzz and the potential barrier as ``near-fuzz" region in analogy with the ``near-horizon" region.

AMPS's firewall argument is based on the experience of an infalling observer with a highly blue-shifted quantum $B$ in her own frame.
The interaction with a quantum $B$ { in isolation} is possible { outside} the potential barrier located at $r_*$. Repeating the argument of AMPS for this case we find that quantum $B$ cannot generically be maximally entangled with a quantum in the near-fuzz region after Page time and, according to our analysis in Section \ref{Qubit}, not even before.
Since quantum $B$ is not close to the horizon it is not sufficiently blue shifted to argue for a firewall.
The more interesting case is when the infalling observer crosses the barrier and enters into the near-fuzz region where she now encounters not one but many quanta. To capture the full dynamics of an observer falling into a fuzzball one must deal with the interaction of the observer with all the quanta in the near-fuzz region and more importantly the ``fuzz'' itself.\footnote{It should be noted that Figure \ref{greybody} should not be taken to mean that there is clear distinction between the fuzz and the high-energy quanta in the near-fuzz region. While we would expect the physics to be well described by classical gravity for $r \gtrsim r_*$ the figure should be taken to mean a smooth transition from highly quantum-stringy states at the location of the fuzz to gas of particles on a gravity background at $r \simeq r_*$.} 
Since AMPS analysis relies on effective field theory all the way up to the (streched) horizon both of these additional interactions are absent. While the finding of a diverging stress tensor is evidence that effective field theory breaks down it does not automatically imply that the strong dynamics of the microstates of quantum gravity leads to a universally dramatic encounter for all types of infalling observers. 
To answer what the experience of an infalling observer is one must understand the nature of the quantum gravity microstates.
While the description of the entire dynamics of an infalling observer with a typical fuzzball might seem a daunting task, after scrambling time it seems possible to make an insightful approximation due to Mathur which we elaborate on in the remainder of this section.

\subsection{Fuzzball complementarity}\label{fuzzballcomplementarity}


In this section we briefly review fuzzball complementarity as developed in~\cite{Mathur:2011wg,Mathur:2010kx,Mathur:2012zp,Mathur:2012dx}. For more recent work see~\cite{Mathur:2012jk,Mathur:2013gua}. Our discussion is in the context of the AdS/CFT correspondence. However, we expect the lesson that will be drawn from this to hold more generally.

We start with a typical state in a CFT.
For {\em sufficiently coarse-grained operators} probing the system we can {\em approximate} a typical state by a thermal density matrix $\rho$ by using the standard technique of going from the micro-canonical to the canonical ensemble.\footnote{As emphasized in the previous subsection this approximation cannot be made for fine-grained operators which probe the energy scale of typical quanta of the state.} The temperature of the approximated thermal density matrix is fixed by the dynamics of the system and the energy of the typical state $\ket{\psi}$. The expectation values of coarse-grained operators $\hat O$ in the typical state get approximated by a thermal average over the energy 
eigenstates
\be
\phantom{a} \langle \psi | \hat O | \psi \rangle \approx Tr(\rho \hat O) = \f{1}{\sum_i e^{- \f{ E_i}{T_H}}} \sum_k e^{-\f{E_k}{T_H} } \phantom{a} \langle E_k | \hat O | E_k \rangle\,. \label{expValTyp}
\ee

Using the thermofield double formalism~\cite{doi:10.1142/S0217979296000817} we can purify the density matrix as defined in \eqref{expValTyp} by addition of states in an auxiliary CFT to obtain the pure state
\be
\ket \Psi = \f{1}{\sqrt{\sum_i e^{- \f{ E_i}{T_H}}}} \sum_k e^{-\f{E_k}{2 T_H}} \ket{ E_{k}}_L \otimes \ket{ E_{k}}_R\,, \label{CFTEntangle}
\ee
where the $\ket{E_k}_R$ denote the energy eigenstates of the original CFT, which we call $CFT_R$, while $\ket{E_k}_L$ are the energy eigenstates of an auxiliary $CFT_L$. The subscripts' interpretation will become clear shortly.
With this we can rewrite \bref{expValTyp} as 
\be
\phantom{a}_R \langle \psi | \hat O_R | \psi \rangle_R \approx \langle \Psi | \hat O_R | \Psi \rangle\,, \label{TwoCFTExpVal}
\ee
where we have put a subscript on the operator and the typical state of Equation  \bref{expValTyp} as well. The expectation value of a coarse-grained operator $\hat O_R$ in a state $\ket{\psi}_R$ can thus be approximately calculated in state $\ket{\Psi}$ that is entangled between two copies of the CFT. 

\begin{figure}[htbp]
\begin{center}
 \includegraphics[scale=.30]{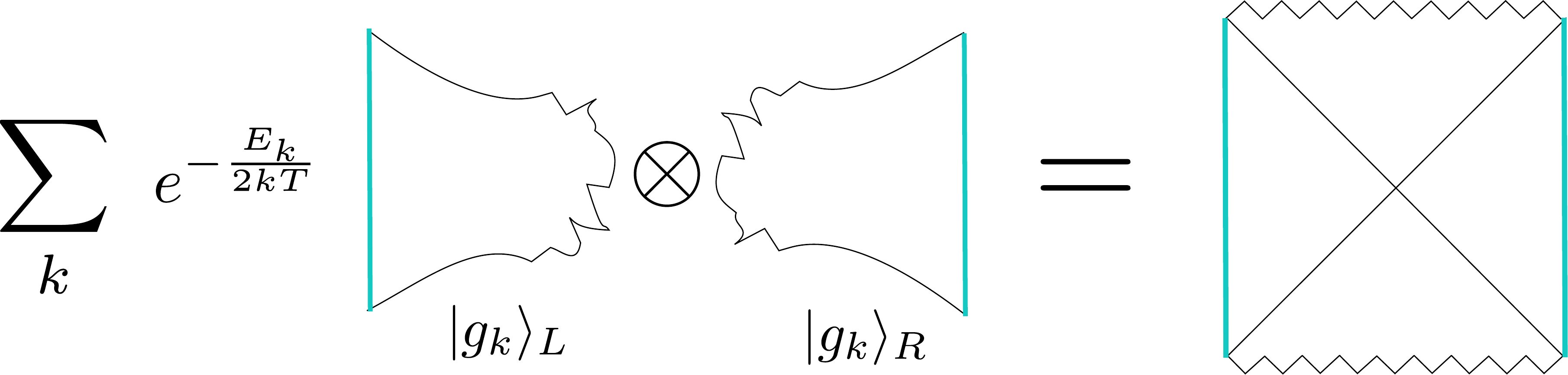}
\caption{The extended AdS Schwarzschild black hole can be understood as the sum over entangled fuzzball solutions $\ket{g_k}_L$ and $\ket{g_k}_R$.}
\label{fn7}
\end{center}
\end{figure}

Maldacena has proposed in \cite{Maldacena:2001kr} that the state \bref{CFTEntangle} is dual to the eternal AdS black hole. In \cite{VanRaamsdonk:2009ar} Van Raamsdonk has recently taken this notion of entanglement of states of CFTs  further to the {\it entanglement of asymptotically gravitational  solutions} further stating his ideas to be very similar to fuzzballs~\cite{Czech:2012be}. Each state $\ket{E_k}$ of one of CFTs should be dual to a fuzzball $\ket{g_k}$. 
While the two CFTs are entangled but not connected, their entanglement results in a description of connected asymptotically $AdS$ solutions.  The Penrose diagram of an eternal AdS black hole can then be understood as an entanglement of fuzzball solutions $\ket{g_k}_L$ and $\ket{g_k}_R$. This is depicted in \mbox{figure \ref{fn7}}.

\begin{figure}[htbp]
\begin{center}
\subfigure[]{
\includegraphics[scale=.35]{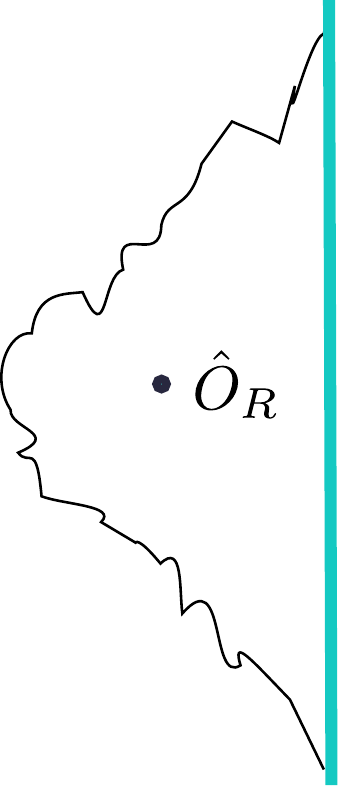}
}
\hspace{3cm}
\subfigure[]{
\includegraphics[scale=.35]{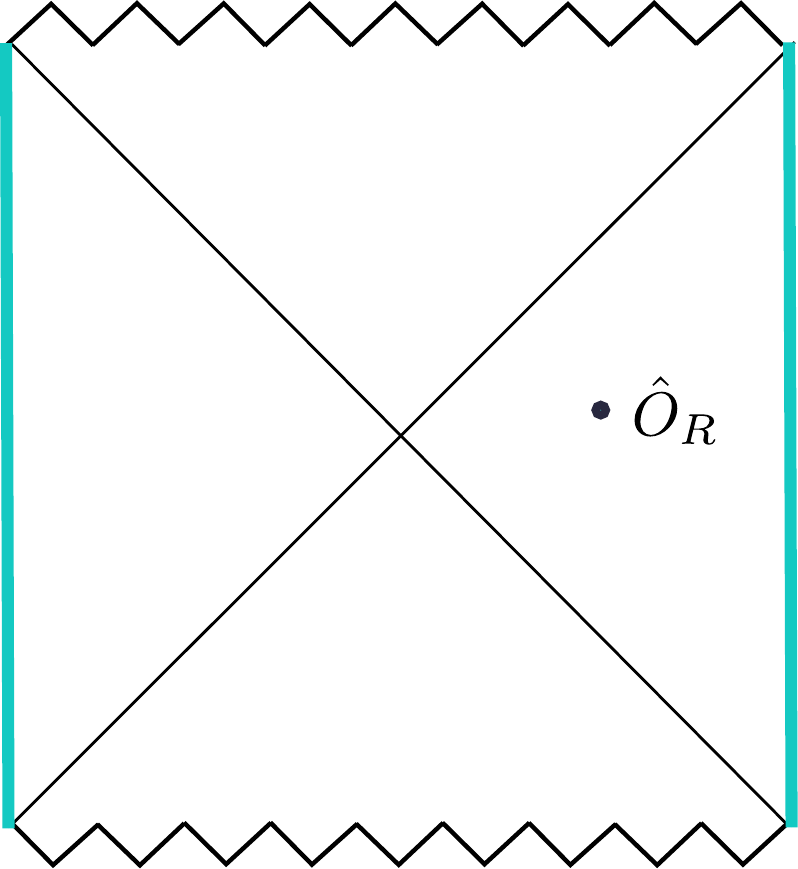}
}
\caption{{(a) Expectation value of an operator $\hat O_R$ in a given fuzzball state $\ket{\psi_g}_R$. (b) For a suitably coarse-grained $\hat O_R$ in a typical $\ket{\psi_g}_R$, this expectation value can be approximately obtained in the eternal AdS geometry.}}
\label{fn5q}
\end{center}
\end{figure}

What are the implications of all this for an observable $\hat O$ measured in a typical  fuzzball $\ket{g}_R$? From the above discussion we see that  we can obtain an approximate answer by inserting the operator in the eternal AdS Schwarzschild solution as depicted in Figure~\ref{fn5q} which is the bulk dual of \bref{TwoCFTExpVal}.

This is not unexpected for time independent operator which corresponds to an observer who is at a constant radial location in the bulk.  But what can we say about the experience of an infalling observer which corresponds to a time dependent operator?  
According to the proposal the above picture of free infall is still valid as long as the infalling observer corresponds to a sufficiently coarse-grained measurement. Since outgoing typical quanta have Killing energy $E \sim T_H$, an incoming quantum of such energy will experience the fine-grained structure of the fuzzball and will thus not evolve in a way describable by infall in a black hole geometry. An incoming high-energy quantum with Killing energy $E \gg T_H$, on the other hand, is insensitive to the details of the fuzzball and may evolve in a way describable by semi-classical evolution in a black hole geometry.

 \begin{figure}[htbp]
\begin{center}
\subfigure[]{
\includegraphics[scale=.35]{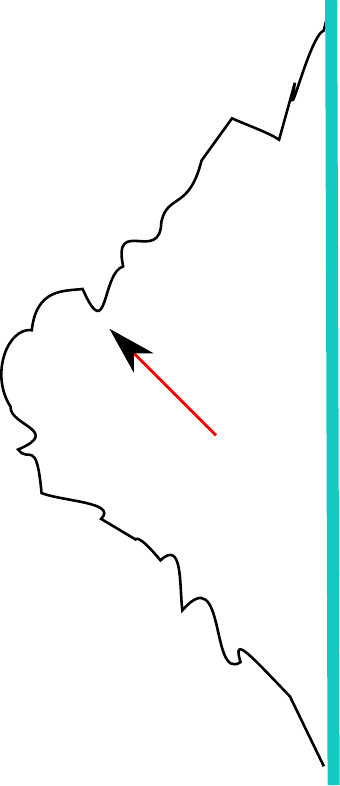}
}
\hspace{3cm}
\subfigure[]{
\includegraphics[scale=.35]{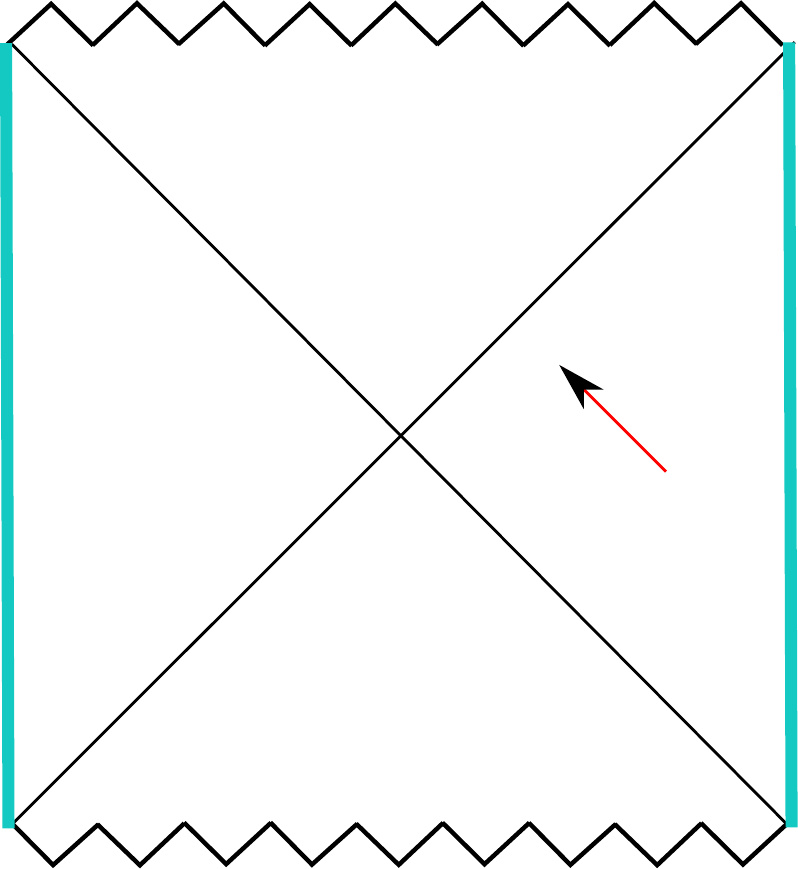}
} \\
\subfigure[]{
\includegraphics[scale=.35]{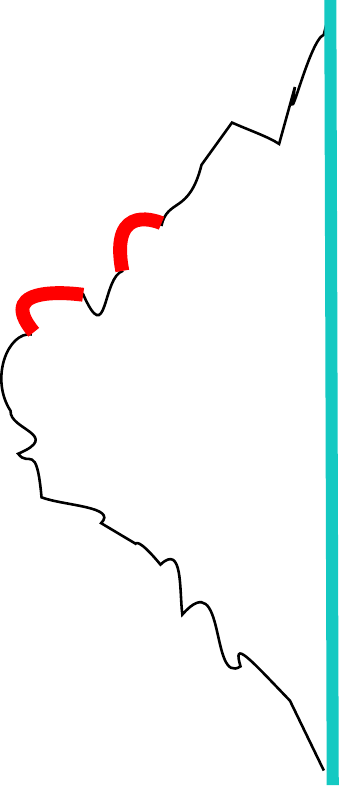}
}
\hspace{3cm}
\subfigure[]{
\includegraphics[scale=.35]{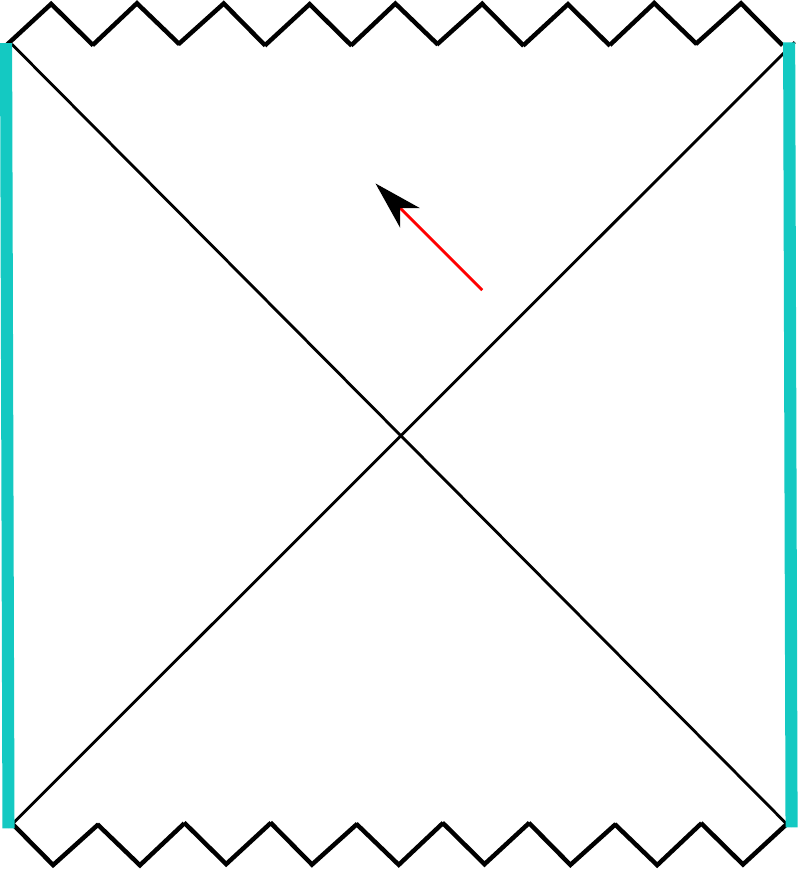}
}
\caption{In (a) and (c) we show the process of infall into a typical fuzzball. Based on the discussion in the main body we expect that a high-energy quanta hitting such a fuzzball will be absorbed by it and excite its collective modes. This process is expected to have an alternate description of infall into a black hole shown in figure (b) and (d). Note that this description is short-lived as is evidenced by the presence of a singularity. This feature is related to the limited phase space coming from the finiteness of the Hilbert space of the fuzzball wave function.}
\label{fuzzballInfall}
\end{center}
\end{figure}

How should we understand this process? Since only the fuzzballs on the right $\ket{g}_R$ are real, where did this extra spacetime beyond the fuzz come from? The infall into the black hole should be viewed as an alternate, auxiliary, description of the, otherwise complicated, interaction of an infalling observer hitting the fuzz and exciting its collective degrees of freedom.\footnote{This is not an unfamiliar situation. A closed string hits a stack of D-branes and yet has a description of falling into AdS spacetime.} However, this  description is short-lived as there is a singularity which the infalling observer eventually hits. This would probably be related to the energy in the collective modes leaking into the thermal modes because of the finite dimensionality of the Hilbert space of fuzzballs.
We have studied the two extreme cases of typical modes and high-energy infalling observers. For observers of intermediate energies we would expect an intermediate picture.

\begin{figure}[htbp]
\begin{center}
\includegraphics[scale=.45]{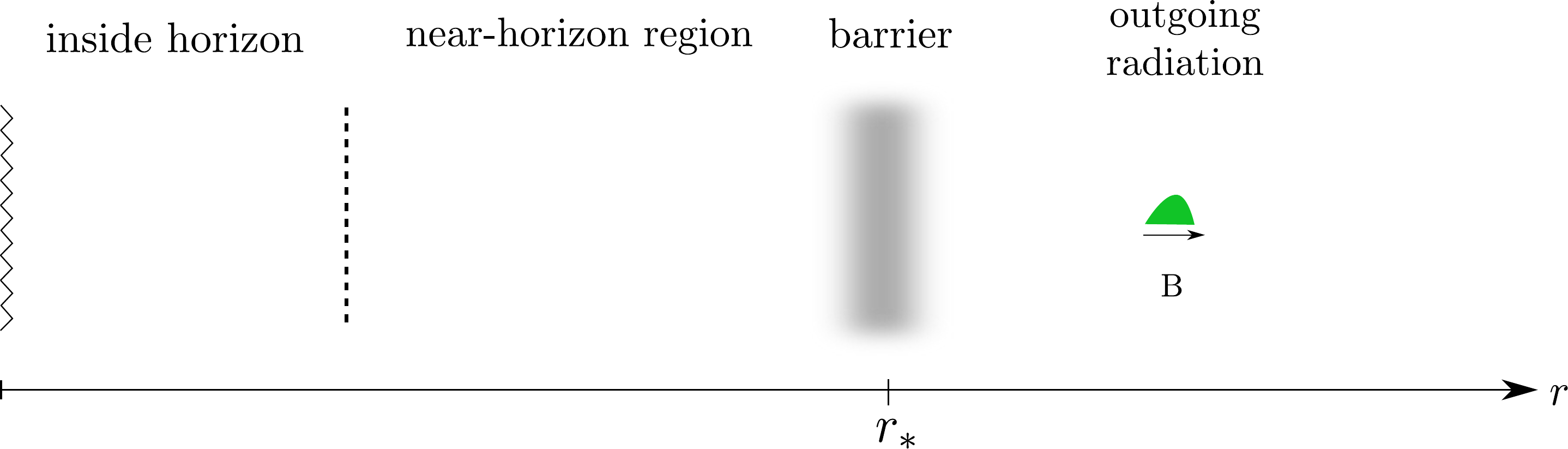}
\caption{The region inside the potential barrier at $r_*$ of Figure \ref{greybody} can be approximated by a black hole geometry in the Hartle-Hawking state. Unitarity requires that such an approximation be valid only for a time scale less that $\sim M$. In other words it is not consistent to patch such geometries together in time for times larger than $\sim M$. A high-energy ($E \gg  T_H$ ) infalling quantum experiences the black hole geometry. Since the time for infall is $\sim M$ it is consistent with the limits on the approximation. However, a low energy quantum ($E \sim T_H$) does not experience a universal infall because it can distinguish between micro-canonical and canonical ensemble.}
\label{greybodyinfalling1dBH}
\end{center}
\end{figure}

In the discussion we used AdS/CFT but recall that according to the arguments in \cite{Maldacena:1996ix} the CFT captures the physics inside the potential barrier.  The decoupling limit is when the potential barrier become infinitely high and for certain solutions this is a consistent limit to take. It is such solutions which have a near-horizon AdS region in the decoupling limit. While for the Schwarzschild black hole there is no decoupling limit, we still expect the general lesson we obtained above in the nearly decoupled case to hold. In other words,  the arguments of the preceding subsection suggest that for sufficiently coarse-grained ($E\gg T_H$) operators we may approximate the region inside the potential barrier in Figure~\ref{greybody} by an eternal Schwarzschild black hole in the Hartle-Hawking state as shown in Figure~\ref{greybodyinfalling1dBH}. A high-energy infalling observer can thus experience a drama-free infall.\footnote{Exact details will depend on whether the observer is modeled by a 
simple Unruh-de Witt detector or something more complicated. This will decide how coarse-grained the operator is.}
For fine-grained ($E \sim T_H$) operators on the other hand we cannot approximate a typical fuzzball by a black hole. Indeed, this is consistent with the idea that typical quanta which carry away information of the system should not have a universal interaction with the system when reflected back towards it as was alluded to in the introduction.

A word of caution is necessary here. The central theme of fuzzball complementarity is an approximation of the microcanonical ensemble by the canonical ensemble. For appropriate coarse-grained observables a typical pure state (a typical fuzzball) can be approximated by a thermal state (a black hole). While a ``zeroed-out" Unruh-de Witt detector tuned to high energies being dropped from just outside the potential barrier at $r_*$ qualifies for this, the same detector when it stays in the near fuzz region at a constant radius does not. Such a detector necessarily has access to a larger part of the system. For instance such a detector staying outside for longer than Page time will experience a drop in the outside radiation entropy. Someone making readings on it will thus conclude that the geometry is not that of a black hole and conclude semi-classical physics is violated at the horizon scale as the black hole geometry leads to an ever increasing entropy. This should not come as a surprise because indeed 
fuzzball proposal does state that the black hole geometry is an approximation.

\subsection{Fuzzball complementarity is not ruled out by AMPS}\label{fuzzballcomplementarityAMPS}

One may think that running the AMPS argument on fuzzballs one would rule out fuzzball complementarity after Page time~\cite{Avery:2013exa,Almheiri:2013hfa}. In fact the new result of this paper is that the interface of the black hole and exterior cannot be in a pure state at any time, particularly even before Page time. While this supports the fuzzball conjecture, one may then run an AMPS like argument to say a mixed state at the interface precludes fuzzball complementarity at all times. This conclusion is however premature and the AMPS argument does not rule out the possibility of fuzzball complementarity as demonstrated in~\cite{Mathur:2013gua} which we briefly review. While we have worked in Planck units in most of this paper, for comparison of scales we briefly restore Planck length in this section.

When a mode of Killing energy $E =  n T$ impinges on a fuzzball the number of possible states grows exponentially~\cite{Mathur:2012jk}
\be
\Delta \mathcal N = e^{4 \pi G M^2}(e^{n}-1).
\ee
If $n \gg 1$, in other words if the infalling quantum has Killing energy $E \gg T$ and we are in the realm of validity of the approximation required for fuzzball complementarity, exponentially more degrees of freedom become available which are not entangled with anything.
Assuming a strong interaction  and  fast scrambling dynamics one may then ``use'' these extra degrees of freedom to create many new (analogues of) $BC$ pairs as an effective description of the complicated interaction.  This argument applies both before and after Page time. The new degrees of freedom may account for the entanglement structure in the auxiliary black hole picture needed for free infall~\cite{MathurTurton, Mathur:2012jk}. 

One may worry that the additional energy the infalling observer carries into the fuzzball requires the stretched horizon to behave in an acausal way in order to avoid the AMPS argument. In particular, the new degrees of freedom are supposed to be created before the impact of the observer. In a sense that is not unexpected. If all $e^S$ fuzzball degrees of freedom are confined to the stretched horizon then the Bekenstein-Hawking formula suggests that the surface of the fuzz has to grow before it can accommodate an infalling quantum. In \cite{Mathur:2013gua} the stretching of the fuzzball surface was computed to move the stretched horizon by
\be
\delta r \sim  \sqrt{n} ~ l_p .
\label{stretch}
\ee
From a general relativity perspective this is an apparently acausal behavior. However, it is dictated by the reasonable assumption that the quantum gravity degrees of freedom live just outside the horizon. Such an effect of quantum gravity shows that general relativity thus ceases to be a good description. 
If general relativity breaks down spacetime concepts like causality lose their meaning. 

If one does want to keep using such concepts then one will find them violated and this is no surprise.
Further, note that while the distance by which the fuzzball surface stretches may be very large for $E \gg T$, according to \eqref{stretch} this is still of the order a Planck distance from the stretched horizon. This seemingly acausal, or non-local, behavior is a lot milder than the non-locality advocated in recent works arguing against fuzzballs (and firewalls).\footnote{See for example~\cite{Papadodimas:2012aq} which advocates non-locality over distances $ l_p^4 M^3$ over which the Hawking radiation is spread out. For solar mass black hole  this is about $10^{77}$ light years.}

\section{Conclusion} \label{Conclusion}

The information paradox and the infall problem have been long-standing puzzles in the understanding of black holes. 
Black hole complementarity was advanced to reconcile unitary black hole evaporation with the common lore of free infall through the horizon of a  black hole. AMPS recently re-examined the self-consistency of BHC. Basically arguing for the contrapositive of Mathur's result from  \cite{Mathur:2009hf}---purity of the final state requires the state at the horizon not to be pure (and therefore not the Unruh vacuum) no later than Page time---AMPS observe that an infalling observer encounters quanta outside the horizon in a highly blue shifted state. They conclude that an infalling observer burns up at the horizon.  
Follow-up papers by Susskind~\cite{Susskind:2012rm,Susskind:2012uw} and Bousso~\cite{Bousso:2012as} largely agree with AMPS' result. 
%

The first part of their argument, that quantum gravity effects kick in at the scale of the horizon, relies on Mathur's theorem~\cite{Mathur:2009hf} which is at the core of the fuzzball proposal.
The implications of such quantum gravity effects for infalling observers have, however, caused quite a bit of controversy.
In this paper we disentangled the arguments regarding preservation of unitarity and those regarding infall and argued that fuzzballs provide a resolution to the information question and a very interesting picture for the infall question.

\subsection*{ I. Unitarity}

In Section \ref{Qubit} we used qubit models to show that:
\begin{enumerate}
 \item Unitary evaporation requires \emph{information of the original state to come out in every step of the evolution}.
 \item There is \emph{no difference between before and after Page time as far as information release is concerned}.\footnote{As noted before, while \emph{decoding} of information may only be possible when having access to more than half of the evaporating system (after Page time), the \emph{release} of information must start with the very first quantum being emitted.}
\end{enumerate}
The recent results of Mathur \cite{Mathur:2009hf} have shown that small corrections to the semi-classical Hawking evaporation cannot ensure purity of final state. This might lead one to think that  the evaporation process may be described by the usual semi-classical Hawking evaporation before Page time as long as after Page time there are large corrections to it. We found that the traditional picture of the state at the horizon being the vacuum state at \emph{any} time of the evolution is inconsistent with unitarity and so modifying the state only after Page time is not enough. In the language of AMPS this would mean that the $BC$ system cannot be in the Unruh vacuum at any point during the evaporation of the system, where $B$ is the outgoing qubit and $C$ is a one qubit worth of the system remaining behind.

Not having the vacuum state at the horizon implies a divergent stress tensor. Diverging energies and densities and singularities, especially in a non-renormalizable theory, should be taken as a sign that new physics is appearing, the accurate description of which is beyond the scope of the theory. The appearance of a divergent stress tensor should thus be taken as evidence that quantum gravity effects kick in not just at the Planck scale but at the scale of the horizon. In fact, Mathur proposed this a long time ago \cite{Mathur:2005zp} to advance the fuzzball conjecture.

According to the fuzzball proposal black holes are a coarse-grained description of the true microstates accounting for the Bekenstein--Hawking entropy which are horizonless and singularity-free solutions of quantum gravity. Being coarse-grained, black holes cannot be used to study processes that require knowledge of the microscopic structure of the state, in particular we have shown in Section \ref{Qubit} that black holes cannot evaporate unitarily. The radiation coming from fuzzballs, however, carries away the information of the state and there is no information \emph{paradox}.

\subsection*{ II. Infall into fuzzballs - The possibility of fuzzball complementarity} 

Elaborating on a recent proposal by Mathur we have argued in Section \ref{Approximate Complementarity} how and when the region inside the potential barrier of a typical fuzzball (Figure \ref{greybody}) can by approximated by the eternal black hole in the Hartle-Hawking state (Figure \ref{greybodyinfalling1dBH}).
The procedure discussed is based on the approximation of a typical fuzzball state by a thermal state.

While the approximating procedure shows \emph{how} it may be possible to get the black hole as a coarse-grained description of fuzzballs it is essential to keep in mind \emph{when} it is applicable:
\begin{itemize}
 \item A thermal state can be a good approximation for \emph{typical} states only.
  \item Infalling quanta of typical energy $(E \sim T_H)$ experience the details of the fuzzballs and so their evolution cannot be described by free infall.
 \item Free infall is only possible for high-energy ($E\gg T_H$) observers represented by \emph{sufficiently coarse-grained operators} probing the typical state.
 \item While we have only discussed the extreme limits of $E\sim T_H$ and $E\gg T_H$, infalling observers with Killing energies in between are expected to experience a continuous variation of dramatic to free infall.
\end{itemize}
The physical mechanism behind fuzzball complementarity is the strong dynamics of the infalling observer with the typical fuzzball state which results in the excitation of collective modes. While a detailed description of this process is presently not feasible we can attempt to answer the infall question by making the above approximation. 

This has revealed a crucial feature of fuzzball complementarity: the \emph{separation of energy scales}. While black hole complementarity and observer complementarity advance the idea of a \emph{frame change}  such that infalling and asymptotic observers may have different experiences, for fuzzball complementarity an operator probing a system will yield different answers (experiences) according to whether it is coarse-grained or fine-grained. While high-energy infalling observers that are insensitive to the details of the microstates (coarse-grained) may experience free infall into a black hole, infalling low-energy observers may be seen as the mirror image of the outgoing radiation quanta (fine-grained) and thus do not have such a drama-free experience.

It would be interesting to understand  the implications of this scale dependent complementarity for some issues tangentially related to black holes. For example a model of brane bound states for early universe was discussed in~\cite{Chowdhury:2006pk,KalyanaRama:2007eb,Bhowmick:2008cq,Bhowmick:2010dd}. In light of this it would be interesting to understand the implications of scale dependent complementarity on resolution of the early universe singularity. Furthermore, it would also be interesting to understand why general relativity is able to capture the coarse-grained solution and the number of microstates this solution is an approximation to. It has indeed been emphasized that Einstein's equations themselves can be obtained from thermodynamics \cite{Jacobson:1995ab} and speculations have been made on the entropic nature of gravity \cite{Verlinde:2010hp}. It would be interesting to understand the implications of scale dependent complementarity for these issues also. The similarities of tunneling of 
infalling shells into fuzzballs with the (opposite of) emergence of spacetime by decoherence \cite{Kiefer} is striking and scale 
dependent complementarity may have something to add in this discussion also.

\section*{Acknowledgments}

BDC would like to thank  Marco Baggio, Jan de Boer, Gilad Lifschytz, Samir D. Mathur, Ashoke Sen, Larus Thorlacius, David Turton, Erik Verlinde and Bram Wouters for discussions. AP would like to thank Iosif Bena, Sheer El-Showk, David Turton and Erik Verlinde for discussions. We would like to thank Steve Giddings, Finn Larsen, Don Marolf, Joe Polchinski, Lenny Susskind, Herman Verlinde for  their many useful comments on a draft of this paper.
The work of BDC is supported by the ERC Advanced Grant 268088-EMERGRAV. AP is supported by DSM CEA-Saclay.
\bibliographystyle{toine}
\bibliography{Papers}

\providecommand{\href}[2]{#2}\begingroup\raggedright\begin{thebibliography}{10}

\bibitem{Gibbons:2009xm}
G.~Gibbons, \emph{{Birkhoff's invariant and Thorne's Hoop Conjecture}},
\href{http://www.arXiv.org/abs/0903.1580}{{\tt 0903.1580}}

\bibitem{Hawking:1974sw}
S.~Hawking, \emph{{Particle Creation by Black Holes}}, Commun.Math.Phys. {\bf
  43} (1975)
199--220

\bibitem{Susskind:1993if}
L.~Susskind, L.~Thorlacius  and J.~Uglum, \emph{{The Stretched horizon and
  black hole complementarity}}, Phys.Rev. {\bf D48} (1993) 3743--3761,
\href{http://www.arXiv.org/abs/hep-th/9306069}{{\tt hep-th/9306069}}

\bibitem{Almheiri:2012rt}
A.~Almheiri, D.~Marolf, J.~Polchinski  and J.~Sully, \emph{{Black Holes:
  Complementarity or Firewalls?}}, JHEP {\bf 1302} (2013) 062,
\href{http://www.arXiv.org/abs/1207.3123}{{\tt 1207.3123}}

\bibitem{Bousso:2012as}
R.~Bousso, \emph{{Complementarity Is Not Enough}},
\href{http://www.arXiv.org/abs/1207.5192}{{\tt 1207.5192}}

\bibitem{Mathur:2012jk}
S.~D. Mathur and D.~Turton, \emph{{Comments on black holes I: The possibility
  of complementarity}},
\href{http://www.arXiv.org/abs/1208.2005}{{\tt 1208.2005}}

\bibitem{Chowdhury:2012vd}
B.~D. Chowdhury and A.~Puhm, \emph{{Is Alice burning or fuzzing?}},
\href{http://www.arXiv.org/abs/1208.2026}{{\tt 1208.2026}}

\bibitem{Susskind:2012rm}
L.~Susskind, \emph{{Singularities, Firewalls, and Complementarity}},
\href{http://www.arXiv.org/abs/1208.3445}{{\tt 1208.3445}}

\bibitem{Banks:2012nn}
T.~Banks and W.~Fischler, \emph{{Holographic Space-Time Does Not Predict
  Firewalls}},
\href{http://www.arXiv.org/abs/1208.4757}{{\tt 1208.4757}}

\bibitem{Ori:2012jx}
A.~Ori, \emph{{Firewall or smooth horizon?}},
\href{http://www.arXiv.org/abs/1208.6480}{{\tt 1208.6480}}

\bibitem{Susskind:2012uw}
L.~Susskind, \emph{{The Transfer of Entanglement: The Case for Firewalls}},
\href{http://www.arXiv.org/abs/1210.2098}{{\tt 1210.2098}}

\bibitem{Hossenfelder:2012mr}
S.~Hossenfelder, \emph{{Comment on the black hole firewall}},
\href{http://www.arXiv.org/abs/1210.5317}{{\tt 1210.5317}}

\bibitem{Page:1993df}
D.~N. Page, \emph{{Average entropy of a subsystem}}, Phys.Rev.Lett. {\bf 71}
  (1993) 1291--1294,
\href{http://www.arXiv.org/abs/gr-qc/9305007}{{\tt gr-qc/9305007}}

\bibitem{Page:1993up}
D.~N. Page, \emph{{Black hole information}},
\href{http://www.arXiv.org/abs/hep-th/9305040}{{\tt hep-th/9305040}}

\bibitem{Mathur:2009hf}
S.~D. Mathur, \emph{{The Information paradox: A Pedagogical introduction}},
  Class.Quant.Grav. {\bf 26} (2009) 224001,
\href{http://www.arXiv.org/abs/0909.1038}{{\tt 0909.1038}}

\bibitem{Mathur:2010kx}
S.~D. Mathur, \emph{{The Information paradox and the infall problem}},
  Class.Quant.Grav. {\bf 28} (2011) 125010,
\href{http://www.arXiv.org/abs/1012.2101}{{\tt 1012.2101}}

\bibitem{Mathur:2011wg}
S.~D. Mathur and C.~J. Plumberg, \emph{{Correlations in Hawking radiation and
  the infall problem}}, JHEP {\bf 1109} (2011) 093,
\href{http://www.arXiv.org/abs/1101.4899}{{\tt 1101.4899}}

\bibitem{Mathur:2011uj}
S.~D. Mathur, \emph{{What the information paradox is {\it not}}},
\href{http://www.arXiv.org/abs/1108.0302}{{\tt 1108.0302}}

\bibitem{Mathur:2012np}
S.~D. Mathur, \emph{{The information paradox: conflicts and resolutions}},
\href{http://www.arXiv.org/abs/1201.2079}{{\tt 1201.2079}}

\bibitem{Giddings:2012dh}
S.~B. Giddings and Y.~Shi, \emph{{Quantum information transfer and models for
  black hole mechanics}},
\href{http://www.arXiv.org/abs/1205.4732}{{\tt 1205.4732}}

\bibitem{Avery:2011nb}
S.~G. Avery, \emph{{Qubit Models of Black Hole Evaporation}},
\href{http://www.arXiv.org/abs/1109.2911}{{\tt 1109.2911}}

\bibitem{Braunstein:2009my}
S.~L. Braunstein, \emph{{Entangled black holes as ciphers of hidden
  information}},
\href{http://www.arXiv.org/abs/0907.1190}{{\tt 0907.1190}}

\bibitem{Mathur:2012zp}
S.~D. Mathur, \emph{{Black Holes and Beyond}},
\href{http://www.arXiv.org/abs/1205.0776}{{\tt 1205.0776}}

\bibitem{Mathur:2012dx}
S.~D. Mathur, \emph{{Black holes and holography}},
\href{http://www.arXiv.org/abs/1207.5431}{{\tt 1207.5431}}

\bibitem{Giddings:2011ks}
S.~B. Giddings, \emph{{Models for unitary black hole disintegration}},
\href{http://www.arXiv.org/abs/1108.2015}{{\tt 1108.2015}}

\bibitem{Giddings:2012bm}
S.~B. Giddings, \emph{{Black holes, quantum information, and unitary
  evolution}}, Phys.Rev. {\bf D85} (2012) 124063,
\href{http://www.arXiv.org/abs/1201.1037}{{\tt 1201.1037}}

\bibitem{Horowitz:2003he}
G.~T. Horowitz and J.~M. Maldacena, \emph{{The Black hole final state}}, JHEP
  {\bf 0402} (2004) 008,
\href{http://www.arXiv.org/abs/hep-th/0310281}{{\tt hep-th/0310281}}

\bibitem{Lowe:1995ac}
D.~A. Lowe, J.~Polchinski, L.~Susskind, L.~Thorlacius  and J.~Uglum,
  \emph{{Black hole complementarity versus locality}}, Phys.Rev. {\bf D52}
  (1995) 6997--7010,
\href{http://www.arXiv.org/abs/hep-th/9506138}{{\tt hep-th/9506138}}

\bibitem{Mathur:2008wi}
S.~D. Mathur, \emph{{What Exactly is the Information Paradox?}}, Lect.Notes
  Phys. {\bf 769} (2009) 3--48,
\href{http://www.arXiv.org/abs/0803.2030}{{\tt 0803.2030}}

\bibitem{Lieb:1973cp}
E.~Lieb and M.~Ruskai, \emph{{Proof of the strong subadditivity of
  quantum-mechanical entropy}}, J.Math.Phys. {\bf 14} (1973)
1938--1941

\bibitem{Araki:1970ba}
H.~Araki and E.~Lieb, \emph{{Entropy inequalities}}, Commun.Math.Phys. {\bf 18}
  (1970)
160--170

\bibitem{PhysRevLett.49.1683}
W.~H. Zurek, \emph{Entropy Evaporated by a Black Hole}, Phys. Rev. Lett. {\bf
  49} (Dec, 1982)
1683--1686

\bibitem{Czech:2011wy}
B.~Czech, K.~Larjo  and M.~Rozali, \emph{{Black Holes as Rubik's Cubes}}, JHEP
  {\bf 1108} (2011) 143,
\href{http://www.arXiv.org/abs/1106.5229}{{\tt 1106.5229}}

\bibitem{nielsen}
M.~A. Nielsen and I.~L. Chuang, {\em Quantum Computation and Quantum
  Information}.
\newblock Cambridge University Press,
2000

\bibitem{Giddings:2009gj}
S.~B. Giddings and R.~A. Porto, \emph{{The Gravitational S-matrix}}, Phys.Rev.
  {\bf D81} (2010) 025002,
\href{http://www.arXiv.org/abs/0908.0004}{{\tt 0908.0004}}

\bibitem{Hayden:2007cs}
P.~Hayden and J.~Preskill, \emph{{Black holes as mirrors: Quantum information
  in random subsystems}}, JHEP {\bf 0709} (2007) 120,
\href{http://www.arXiv.org/abs/0708.4025}{{\tt 0708.4025}}

\bibitem{Sekino:2008he}
Y.~Sekino and L.~Susskind, \emph{{Fast Scramblers}}, JHEP {\bf 0810} (2008)
  065,
\href{http://www.arXiv.org/abs/0808.2096}{{\tt 0808.2096}}

\bibitem{PhysRev.82.664}
J.~Schwinger, \emph{On Gauge Invariance and Vacuum Polarization}, Phys. Rev.
  {\bf 82} (Jun, 1951)
664--679

\bibitem{PhysRev.57.315}
L.~I. Schiff, H.~Snyder  and J.~Weinberg, \emph{On The Existence of Stationary
  States of the Mesotron Field}, Phys. Rev. {\bf 57} (Feb, 1940)
315--318

\bibitem{Chowdhury:2008bd}
B.~D. Chowdhury and S.~D. Mathur, \emph{{Pair creation in non-extremal fuzzball
  geometries}}, Class. Quant. Grav. {\bf 25} (2008) 225021,
\href{http://www.arXiv.org/abs/0806.2309}{{\tt 0806.2309}}

\bibitem{Giddings:2012gc}
S.~B. Giddings, \emph{{Nonviolent nonlocality}},
\href{http://www.arXiv.org/abs/1211.7070}{{\tt 1211.7070}}

\bibitem{Giddings:2013kcj}
S.~B. Giddings, \emph{{Nonviolent information transfer from black holes: a
  field theory parameterization}},
\href{http://www.arXiv.org/abs/1302.2613}{{\tt 1302.2613}}

\bibitem{Chowdhury:2013mka}
B.~D. Chowdhury, \emph{{Cool horizons lead to information loss}},
\href{http://www.arXiv.org/abs/1307.5915}{{\tt 1307.5915}}

\bibitem{Mathur:2005zp}
S.~D. Mathur, \emph{{The Fuzzball proposal for black holes: An Elementary
  review}}, Fortsch.Phys. {\bf 53} (2005) 793--827,
\href{http://www.arXiv.org/abs/hep-th/0502050}{{\tt hep-th/0502050}}

\bibitem{Bena:2004de}
I.~Bena and N.~P. Warner, \emph{{One ring to rule them all ... and in the
  darkness bind them?}}, Adv.Theor.Math.Phys. {\bf 9} (2005) 667--701,
\href{http://www.arXiv.org/abs/hep-th/0408106}{{\tt hep-th/0408106}}

\bibitem{Skenderis:2008qn}
K.~Skenderis and M.~Taylor, \emph{{The fuzzball proposal for black holes}},
  Phys.Rept. {\bf 467} (2008) 117--171,
\href{http://www.arXiv.org/abs/0804.0552}{{\tt 0804.0552}}

\bibitem{Balasubramanian:2008da}
V.~Balasubramanian, J.~de~Boer, S.~El-Showk  and I.~Messamah, \emph{{Black
  Holes as Effective Geometries}}, Class.Quant.Grav. {\bf 25} (2008) 214004,
\href{http://www.arXiv.org/abs/0811.0263}{{\tt 0811.0263}}

\bibitem{Chowdhury:2010ct}
B.~D. Chowdhury and A.~Virmani, \emph{{Modave Lectures on Fuzzballs and
  Emission from the D1-D5 System}},
\href{http://www.arXiv.org/abs/1001.1444}{{\tt 1001.1444}}

\bibitem{Chowdhury:2007jx}
B.~D. Chowdhury and S.~D. Mathur, \emph{{Radiation from the non-extremal
  fuzzball}}, Class. Quant. Grav. {\bf 25} (2008) 135005,
\href{http://www.arXiv.org/abs/0711.4817}{{\tt 0711.4817}}

\bibitem{Avery:2010hs}
S.~G. Avery, B.~D. Chowdhury  and S.~D. Mathur, \emph{{Excitations in the
  deformed D1D5 CFT}}, JHEP {\bf 06} (2010) 032,
\href{http://www.arXiv.org/abs/1003.2746}{{\tt 1003.2746}}

\bibitem{Lunin:2001dt}
O.~Lunin and S.~D. Mathur, \emph{{The Slowly rotating near extremal D1 - D5
  system as a `hot tube'}}, Nucl.Phys. {\bf B615} (2001) 285--312,
\href{http://www.arXiv.org/abs/hep-th/0107113}{{\tt hep-th/0107113}}

\bibitem{Giusto:2004ip}
S.~Giusto, S.~D. Mathur  and A.~Saxena, \emph{{3-charge geometries and their
  CFT duals}}, Nucl.Phys. {\bf B710} (2005) 425--463,
\href{http://www.arXiv.org/abs/hep-th/0406103}{{\tt hep-th/0406103}}

\bibitem{Bena:2012zi}
I.~Bena, A.~Puhm  and B.~Vercnocke, \emph{{Non-extremal Black Hole Microstates:
  Fuzzballs of Fire or Fuzzballs of Fuzz ?}},
\href{http://www.arXiv.org/abs/1208.3468}{{\tt 1208.3468}}

\bibitem{Unruh:1983ms}
W.~G. Unruh and R.~M. Wald, \emph{{What happens when an accelerating observer
  detects a Rindler particle}}, Phys.Rev. {\bf D29} (1984)
1047--1056

\bibitem{Mathur:2013gua}
S.~D. Mathur and D.~Turton, \emph{{The flaw in the firewall argument}},
\href{http://www.arXiv.org/abs/1306.5488}{{\tt 1306.5488}}

\bibitem{doi:10.1142/S0217979296000817}
Y.~TAKAHASHI and H.~UMEZAWA, \emph{THERMO FIELD DYNAMICS}, International
  Journal of Modern Physics B {\bf 10} (1996), no.~13n14, 1755--1805,
\href{http://www.arXiv.org/abs/http://www.worldscientific.com/doi/pdf/10.1142/S0217979296000817}{{\tt
  http://www.worldscientific.com/doi/pdf/10.1142/S0217979296000817}}

\bibitem{Maldacena:2001kr}
J.~M. Maldacena, \emph{{Eternal black holes in anti-de Sitter}}, JHEP {\bf
  0304} (2003) 021,
\href{http://www.arXiv.org/abs/hep-th/0106112}{{\tt hep-th/0106112}}

\bibitem{VanRaamsdonk:2009ar}
M.~Van~Raamsdonk, \emph{{Comments on quantum gravity and entanglement}},
\href{http://www.arXiv.org/abs/0907.2939}{{\tt 0907.2939}}

\bibitem{Czech:2012be}
B.~Czech, J.~L. Karczmarek, F.~Nogueira  and M.~Van~Raamsdonk, \emph{{Rindler
  Quantum Gravity}},
\href{http://www.arXiv.org/abs/1206.1323}{{\tt 1206.1323}}

\bibitem{Maldacena:1996ix}
J.~M. Maldacena and A.~Strominger, \emph{{Black hole grey body factors and
  d-brane spectroscopy}}, Phys.Rev. {\bf D55} (1997) 861--870,
\href{http://www.arXiv.org/abs/hep-th/9609026}{{\tt hep-th/9609026}}

\bibitem{Avery:2013exa}
S.~G. Avery and B.~D. Chowdhury, \emph{{Firewalls in AdS/CFT}},
\href{http://www.arXiv.org/abs/1302.5428}{{\tt 1302.5428}}

\bibitem{Almheiri:2013hfa}
A.~Almheiri, D.~Marolf, J.~Polchinski, D.~Stanford  and J.~Sully, \emph{{An
  Apologia for Firewalls}},
\href{http://www.arXiv.org/abs/1304.6483}{{\tt 1304.6483}}

\bibitem{MathurTurton}
S.~D. Mathur and D.~Turton,
\emph{Private communications},

\bibitem{Papadodimas:2012aq}
K.~Papadodimas and S.~Raju, \emph{{An Infalling Observer in AdS/CFT}},
\href{http://www.arXiv.org/abs/1211.6767}{{\tt 1211.6767}}

\bibitem{Chowdhury:2006pk}
B.~D. Chowdhury and S.~D. Mathur, \emph{{Fractional brane state in the early
  universe}}, Class. Quant. Grav. {\bf 24} (2007) 2689--2720,
\href{http://www.arXiv.org/abs/hep-th/0611330}{{\tt hep-th/0611330}}

\bibitem{KalyanaRama:2007eb}
S.~Kalyana~Rama, \emph{{Entropy of anisotropic universe and fractional
  branes}}, Gen.Rel.Grav. {\bf 39} (2007) 1773--1788,
\href{http://www.arXiv.org/abs/hep-th/0702202}{{\tt hep-th/0702202}}

\bibitem{Bhowmick:2008cq}
S.~Bhowmick, S.~Digal  and S.~K. Rama, \emph{{Stabilisation of Seven (Toroidal)
  Directions and Expansion of the remaining Three in an M theoretic Early
  Universe Model}}, Phys.Rev. {\bf D79} (2009) 101901,
\href{http://www.arXiv.org/abs/0810.4049}{{\tt 0810.4049}}

\bibitem{Bhowmick:2010dd}
S.~Bhowmick and S.~K. Rama, \emph{{10 + 1 to 3 + 1 in an Early Universe with
  mutually BPS Intersecting Branes}}, Phys.Rev. {\bf D82} (2010) 083526,
\href{http://www.arXiv.org/abs/1007.0205}{{\tt 1007.0205}}

\bibitem{Jacobson:1995ab}
T.~Jacobson, \emph{{Thermodynamics of space-time: The Einstein equation of
  state}}, Phys.Rev.Lett. {\bf 75} (1995) 1260--1263,
\href{http://www.arXiv.org/abs/gr-qc/9504004}{{\tt gr-qc/9504004}}

\bibitem{Verlinde:2010hp}
E.~P. Verlinde, \emph{{On the Origin of Gravity and the Laws of Newton}}, JHEP
  {\bf 1104} (2011) 029,
\href{http://www.arXiv.org/abs/1001.0785}{{\tt 1001.0785}}

\bibitem{Kiefer}
C.~Kiefer, \emph{Decoherence in Situations Involving the Gravitational Field},
  in {\em Decoherence: Theoretical, Experimental, and Conceptual Problems},
  P.~Blanchard, E.~Joos, D.~Giulini, C.~Kiefer  and I.-O. Stamatescu, eds.,
  vol.~538 of {\em Lecture Notes in Physics}, pp.~101--112.
\newblock Springer Berlin / Heidelberg,
2000.
\newblock

\end{thebibliography}\endgroup

\end{document}